\documentclass[twocolumn, switch]{article} 

\usepackage{preprint}
\usepackage{algorithm}
\usepackage{algorithmic}
\usepackage{amsmath, amsthm, amssymb, amsfonts}

\usepackage{appendix}
\usepackage{natbib}

\usepackage[utf8]{inputenc}	
\usepackage[T1]{fontenc}	
\usepackage{xcolor}		
\usepackage[colorlinks = true,
            linkcolor = purple,
            urlcolor  = blue,
            citecolor = cyan,
            anchorcolor = black]{hyperref}	
\usepackage{booktabs} 		
\usepackage{nicefrac}		
\usepackage{microtype}		
\usepackage{lineno}		
\usepackage{float}			

\usepackage{lipsum}		
\usepackage{enumitem}
\usepackage{newfloat}
\DeclareFloatingEnvironment[name={Supplementary Figure}]{suppfigure}
\usepackage{sidecap}
\sidecaptionvpos{figure}{c}
\usepackage{subcaption}
\usepackage{graphicx}
\usepackage{titlesec}
\titlespacing\section{0pt}{12pt plus 3pt minus 3pt}{1pt plus 1pt minus 1pt}
\titlespacing\subsection{0pt}{10pt plus 3pt minus 3pt}{1pt plus 1pt minus 1pt}
\titlespacing\subsubsection{0pt}{8pt plus 3pt minus 3pt}{1pt plus 1pt minus 1pt}

\usepackage[table]{xcolor}
\usepackage{soul}
\usepackage{pifont}
\newcommand{\cmark}{\ding{51}} 
\newcommand{\xmark}{\ding{55}}
\usepackage{multirow}

\title{ASK: Adaptive Self-improving Knowledge Framework for Audio Text Retrieval}

\usepackage{eso-pic}
\usepackage{tikz}
\usepackage{xcolor}
\PassOptionsToPackage{colorlinks=true,linkcolor=gray,urlcolor=gray}{hyperref}

\usepackage{titling}
\usepackage{footmisc}
\newcommand{\Author}[2]{\textbf{#1}\textsuperscript{#2}}

\author{
  \Author{Siyuan Fu}{1}\textsuperscript{*} \and
  \Author{Xuchen Guo}{1}\textsuperscript{*} \and
  \Author{Mingjun Liu}{1}\textsuperscript{*} \and
  \Author{Hongxiang Li}{2} \and
  \Author{Boyin Tan} {3} \and
  \Author{Gongxi Zhu}{5}\and  
  \Author{Xianwei Zhuang}{4}\and
  \Author{Jinghan Ru}{4}\and
  \Author{Yuxin Xie}{4}\textsuperscript{$\dagger$}\and
  \Author{Yuguo Yin}{4} \and
}

\date{%
  \textsuperscript{1}University of Electronic Science and Technology of China\\
  \textsuperscript{2} Hong Kong University of Science and Technology\\
  \textsuperscript{3}Mohamed Bin Zayed University of Artificial Intelligence\\
  \textsuperscript{4}Peking University\\
  \textsuperscript{5}Tsinghua University\\
  \texttt{siyuanfu05@gmail.com, yuxinxie2001@gmail.com} \\
}

\begin{document}

\twocolumn[ 
  \begin{@twocolumnfalse} 

\maketitle

\thispagestyle{empty}

\begin{abstract}
The dominant paradigm for Audio-Text Retrieval (ATR) relies on dual-encoder architectures optimized via mini-batch contrastive learning. However, restricting optimization to local in-batch samples creates a fundamental limitation we term the Gradient Locality Bottleneck (GLB), which prevents the resolution of acoustic ambiguities and hinders the learning of rare long-tail concepts. While external knowledge injection can break this bottleneck, it often triggers a problem called Representation-Drift Mismatch (RDM), where a static knowledge base becomes misaligned with evolving encoders, degrading guidance into noise. To address these intertwined challenges, we propose the \textbf{A}daptive \textbf{S}elf-improving \textbf{K}nowledge (\textbf{ASK}) framework. ASK breaks the GLB via multi-grained knowledge injection and mitigates RDM through a dynamic refinement strategy that synchronizes the knowledge base with the model. Additionally, an adaptive reliability weighting scheme is employed to filter retrieval noise based on cross-modal consistency. Extensive experiments across multiple benchmarks demonstrate that ASK consistently achieves new state-of-the-art performance across various backbones. 
\end{abstract}
\vspace{0.35cm}

  \end{@twocolumnfalse} 
] 

\vspace{-0.5em}
{\renewcommand\thefootnote{\fnsymbol{footnote}}%
 \footnotetext[1]{ Equal contribution.}
 \footnotetext[2]{ Corresponding author.}
 
 }


\section{Introduction}
\label{sec:introduction}

\begin{figure}[htbp]
\centering
  \includegraphics[width=\columnwidth]{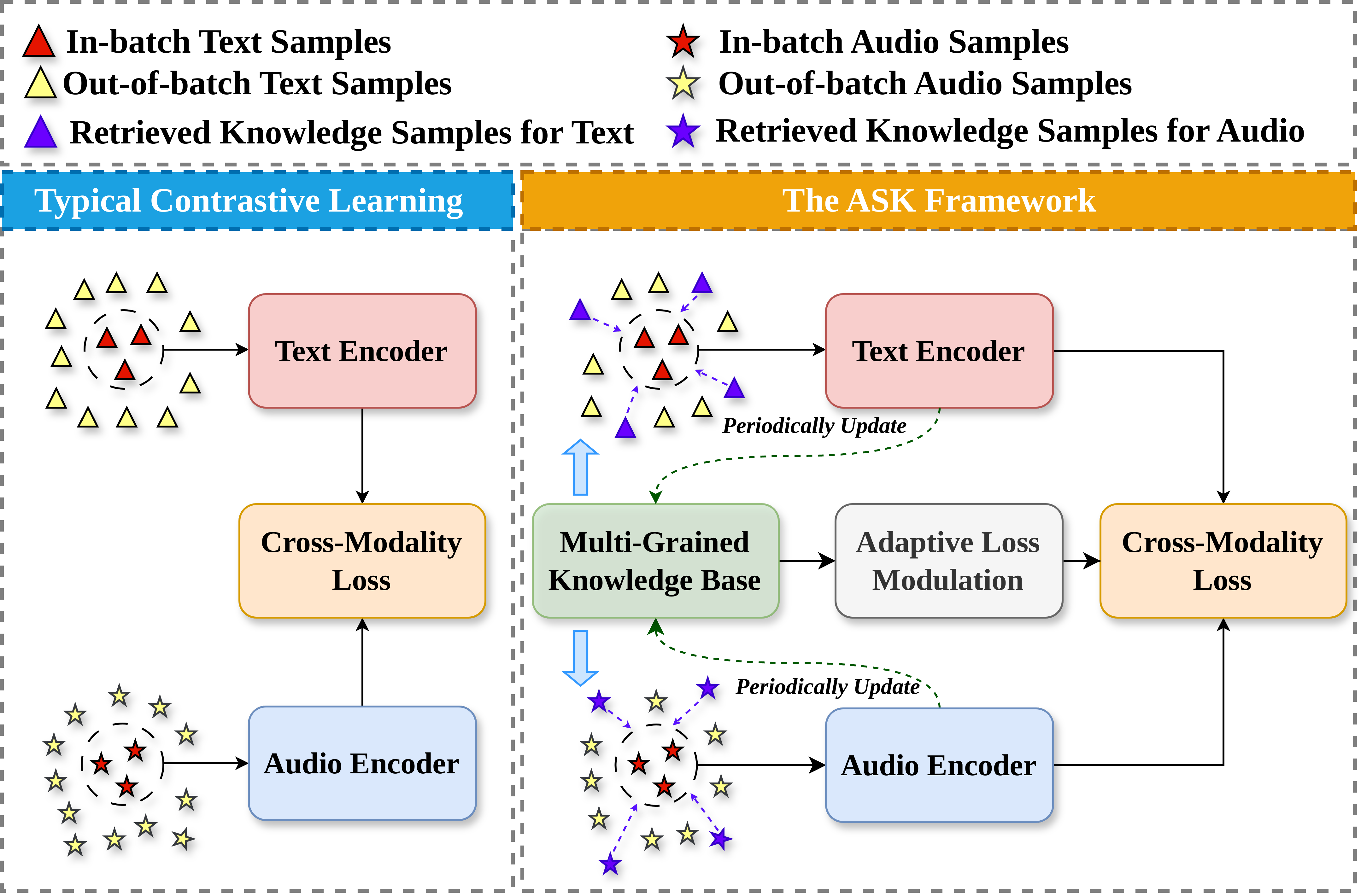}
  \caption{Comparison between typical contrastive learning (left) and our proposed \textbf{ASK} framework (right) with a periodically updated knowledge base and an adaptive loss modulation module.}
  \label{fig:intro}
\end{figure}

Audio-Text Retrieval (ATR) aims to establish a shared embedding space where acoustically and semantically corresponding audio and text pairs are aligned \citep{mei2022metric, yan2024bridging}. The dominant paradigm currently relies on dual-encoder architectures optimized via contrastive learning objectives, such as the NT-Xent loss \citep{chen2020simple}. As illustrated in Figure~\ref{fig:intro} (left), this approach refines representations by maximizing the similarity between matched pairs while contrasting them against other samples within the same mini-batch. While effective, this in-batch mechanism implicitly restricts the optimization landscape to the local context.

The limitations of such batch-constrained optimization have been extensively discussed in broader representation learning literature, particularly regarding the necessity of large negative pools \citep{he2020momentum, xiong2020approximate}. Building on these observations, we define this phenomenon in the context of ATR as the Gradient Locality Bottleneck (GLB). This term characterizes the specific difficulty of resolving fine-grained acoustic details that are often semantically sparse or ambiguous when the model is confined to a limited sample space. Lacking access to a broader global semantic context, the model struggles to form robust decision boundaries for long-tail events, leaving a significant portion of the dataset's semantic potential untapped.

To break this bottleneck, one effective approach is to augment training with an external knowledge base \citep{dwibedi2021little, khandelwal2019generalization, guu2020retrieval}. However, utilizing external memory introduces synchronization latency between the evolving model and the stored representations \citep{xiong2020approximate}.
 We formalize this temporal discrepancy in our continuous training framework as the Representation-Drift Mismatch (RDM). RDM describes the inevitable lag where audio and text encoders update rapidly while the external knowledge base remains static or updates slowly. Consequently, the retrieved knowledge risks degrading from a source of semantic guidance into representational noise, potentially destabilizing the training process.

To systematically address these formally defined challenges, we propose the \textbf{A}daptive \textbf{S}elf-improving \textbf{K}nowledge (ASK) framework. Uniquely designed as a model-agnostic enhancement, ASK can be seamlessly integrated into various backbones. As shown in Figure~\ref{fig:intro} (right), it serves as a holistic solution to inject information from a dynamically maintained multi-grained knowledge base. To handle the inherent noise and ambiguity described by the GLB, we introduce an adaptive reliability mechanism that modulates the learning signal based on cross-modal consistency. Simultaneously, to resolve the RDM, ASK employs a dynamic refinement strategy that periodically updates the knowledge base, ensuring that the external guidance co-evolves with the model.

This synergistic design enables ASK to effectively leverage global semantic structures while maintaining training stability, independent of the specific underlying architecture. Extensive experiments across multiple datasets and backbone models demonstrate that our approach consistently achieves state-of-the-art performance, validating the universality and effectiveness of our framework in handling global optimization challenges.

Our main contributions are summarized as follows:
\begin{itemize}[leftmargin=*]
    \item We identify and define two fundamental challenges in Audio-Text Retrieval: the Gradient Locality Bottleneck (GLB), and the Representation-Drift Mismatch (RDM).
    \item We propose the Adaptive Self-improving Knowledge (ASK) framework, a model-agnostic solution. It breaks the GLB via multi-grained knowledge injection and systematically mitigates RDM through dynamic knowledge refinement. Furthermore, we introduce a novel adaptive reliability weighting scheme to explicitly filter retrieval noise based on cross-modal consistency.
    \item Extensive experiments across multiple benchmarks demonstrate that ASK consistently achieves new state-of-the-art performance across diverse global and local interaction architectures. Comprehensive ablation and zero-shot studies further validate the robustness of our framework.
\end{itemize}

\section{Related Work}
\subsection{Feature Representations}
Feature representation serves as the cornerstone of audio-text retrieval. Early Audio-Text Retrieval systems relied on pairing handcrafted acoustic features like MFCCs \citep{MFCCs} with static word embeddings such as Word2Vec \citep{Word2Vec}. The advent of deep learning has led to the adoption of powerful, pre-trained unimodal encoders. Text representations are now predominantly extracted from large language models like BERT \citep{devlin2019bert}, while audio features are derived from deep models pre-trained on large-scale audio datasets, such as PANNs \citep{kong2020panns} and AST \citep{ast}. More recently, the field has shifted towards large-scale cross-modal pre-training. Models like CLAP \citep{clap, audioclip} leverage contrastive learning on vast audio-text datasets to directly learn a shared embedding space, significantly enhancing zero-shot capabilities. Our work builds upon these advanced encoders, proposing a novel mechanism to further enhance their representations during downstream fine-tuning.

\subsection{Cross-Modal Interaction and Alignment}
Cross-modal interaction is key to achieving semantic alignment in ATR. Early and prevalent approaches perform this at a global, sentence-level, using contrastive learning to align the final embeddings of entire audio clips and text descriptions \citep{clip, wav2clip, mei2022metric}. To capture more fine-grained relationships, recent works have focused on local, token-level interactions. These methods typically employ attention mechanisms or cross-modal Transformers to model correspondences between audio frames and text tokens \citep{scan, lu2019vilbert, xie2024gpa, yin2025atri}. Our ASK framework is orthogonal to these design choices; it operates on the representations themselves and can be seamlessly integrated with both global and local interaction architectures.

\subsection{Retrieval-Augmented Contrastive Learning}

The challenges formalized as GLB and RDM parallel fundamental bottlenecks in computer vision and information retrieval. To overcome limited in-batch negatives, frameworks like MoCo \cite{he2020momentum} and RocketQA \cite{qu2021rocketqa} expand the contrastive denominator via memory queues, while NNCLR \cite{dwibedi2021little} utilizes external support sets. Furthermore, ANCE \cite{xiong2020approximate} proposes asynchronous updates to mitigate stale indices. However, these classical paradigms rely on unimodal queues or strict hard negative mining, making them highly susceptible to acoustic confusion in ATR. Unlike them, ASK structurally breaks the GLB by directly injecting out-of-batch knowledge into representations. Furthermore, to combat the subsequent representation drift, ASK introduces an adaptive reliability weighting scheme that explicitly evaluates cross-modal consistency to filter out retrieval noise.

\section{Problem Formulation and Analysis}

\subsection{Preliminaries}
\label{sec:preliminaries}

In a standard Audio-Text Retrieval framework, a dual-encoder architecture, comprising an audio encoder $f_\theta(\cdot)$ and a text encoder $g_\phi(\cdot)$, maps an audio-text pair $(a_i, t_i)$ to L2-normalized embeddings $u_i$ and $v_i$. The encoders are optimized via a symmetric NT-Xent loss \citep{chen2020simple} over a mini-batch $B$. For a single view, the loss is:
\begin{equation}
\label{eq:ntxent_loss}
\mathcal{L}_{i} = -\log \frac{\exp(u_i^\top v_i / \tau)}{\sum_{v_j \in B} \exp(u_i^\top v_j / \tau)}, 
\end{equation}
where $\tau$ is a temperature hyperparameter. Crucially, as shown in Eq.~\ref{eq:ntxent_loss}, the contrastive denominator is computed exclusively over samples within the mini-batch $B$. This inherent structural confinement is the direct cause of the bottleneck we analyze next.

\subsection{The Gradient Locality Bottleneck}
\label{sec:glb}

The batch-centric nature of standard contrastive objectives creates a fundamental limitation in Audio-Text Retrieval. To formalize this, we introduce the concept of \textbf{Out-of-Batch Influence (OBI)}, which measures the gradient contribution from data points outside the current mini-batch to the optimization process. Let $\mathcal{D}$ denote the entire training dataset and $B \subset \mathcal{D}$ represent a specific mini-batch. For a batch loss $\mathcal{L}_B$, the OBI is defined as the expected gradient norm with respect to all out-of-batch embeddings:
\begin{equation}
\label{eq:obi}
\text{OBI}(\mathcal{L}_B) = \mathbb{E}_{k \in \mathcal{D} \setminus B} \left[ \left\| \frac{\partial \mathcal{L}_B}{\partial u_k} \right\|_2 + \left\| \frac{\partial \mathcal{L}_B}{\partial v_k} \right\|_2 \right],
\end{equation}
where $u_k$ and $v_k$ are the audio and text embeddings of an out-of-batch sample $k \in \mathcal{D} \setminus B$. 

We argue that a training paradigm suffers from a \textbf{Gradient Locality Bottleneck (GLB)} if its OBI is identically zero, indicating that no gradient flow exists from out-of-batch data to guide the current optimization step. In a standard ATR framework using the symmetric NT-Xent loss (Eq.~\ref{eq:ntxent_loss}), the objective $\mathcal{L}_B$ is formulated exclusively as a function of the in-batch embeddings $\{u_j, v_j\}_{j \in B}$. Consequently, the partial derivatives with respect to any out-of-batch embedding $u_k$ or $v_k$ (where $k \notin B$) are necessarily zero:
\begin{equation}
\forall k \in \mathcal{D} \setminus B: \quad \frac{\partial \mathcal{L}_B}{\partial u_k} = \mathbf{0}, \quad \frac{\partial \mathcal{L}_B}{\partial v_k} = \mathbf{0}.
\end{equation}
This directly results in $\text{OBI}(\mathcal{L}_B) = 0$, proving that standard ATR optimization is strictly constrained by the GLB and cannot leverage the vast semantic information present in out-of-batch data. 

This structural confinement prevents the model from leveraging the vast semantic knowledge present in the majority of the dataset. This leads to two critical failures: (1) Semantic Ambiguity and Acoustic Hallucination, where the lack of diverse out-of-batch context prevents the model from learning fine-grained acoustic distinctions between similar but semantically distinct events; and (2) Long-tail Concept Collapse, as the reliance on limited in-batch negatives hinders the formation of robust decision boundaries for rare events, causing the model to default to common concepts with shared tonal properties.

\subsection{The Representation Drift Mismatch}
\label{sec:rdm}
A direct approach to break the GLB is to perform knowledge injection, where out-of-batch samples are retrieved and fused with the current samples. By employing a simple feature fusion such as $u'_i = (1-\rho)u_i + \rho \mathcal{K}$, where $u_i$ is the embedding of the current sample, $\mathcal{K}$ is the retrieved out-of-batch knowledge and $\rho$ represents the injection ratio, we establish a non-zero gradient pathway from the out-of-batch data to the model parameters. This ensures that the Out-of-Batch Influence (OBI) is no longer zero, effectively breaking the structural confinement of standard contrastive learning.

While knowledge injection breaks the GLB by introducing out-of-batch samples $\mathcal{D} \setminus B$, it inherently introduces a critical challenge: \textbf{Representation Drift Mismatch (RDM)}. Since the encoders $f_{\theta_t}$ and $g_{\phi_t}$ are non-stationary and evolve during optimization, a static knowledge base constructed at step $t_k$ becomes progressively misaligned with the current model state at step $t$ ($t > t_k$). To formalize this, taking the audio modality as an example, we define RDM as the expected Kullback-Leibler (KL) divergence \cite{kullback1951information} between the ideal neighborhood distribution $P_{\text{ideal}}$ and the actual distribution $P_{\text{actual}}$:
\begin{equation}
\begin{aligned}
\label{eq:rdm_definition}
\text{RDM}(t, t_k) &= \mathbb{E}_{a_i \in \mathcal{D}} \left[ D_{\text{KL}} \left( P_{\text{ideal}}(\cdot|i) \,||\, P_{\text{actual}}(\cdot|i) \right) \right],\\
P_{\text{ideal}}(j|i) &\propto \exp(\text{sim}(f_{\theta_t}(a_i), f_{\theta_t}(a_j))),\\
P_{\text{actual}}(j|i) &\propto \exp(\text{sim}(f_{\theta_t}(a_i), f_{\theta_{t_k}}(a_j))),
\end{aligned}
\end{equation}
where $a_i$ denotes the current query sample, the distributions are normalized over all knowledge samples indexed by $j$, $\theta_t$ and $\theta_{t_k}$ represent the model parameters at the current step $t$ and the knowledge snapshot step $t_k$ respectively, and $\text{sim}(\cdot, \cdot)$ is the dot product similarity.

Formally, the knowledge vector $\mathcal{K}$ is computed as the expected representation over the neighborhood distribution: $\mathcal{K} = \sum_j P(j) z_j$, where $z_j$ denotes the embedding of the $j$-th knowledge sample. As the training progresses, RDM accumulates and corrupts the optimization objective by inducing a deviation in the fused knowledge vectors $\mathcal{K}$, $\Delta\mathcal{K} = \mathcal{K}_{\text{actual}} - \mathcal{K}_{\text{ideal}}$.

The impact of this drift on training stability can be quantified via the gradient deviation: $\Delta \nabla = \nabla_{\theta_t} \mathcal{L}_{\text{actual}} - \nabla_{\theta_t} \mathcal{L}_{\text{ideal}}$. Specifically, consider a simplified loss: $\mathcal{L} = \mathcal{L}_{\text{main}}(u_i, u'_i)$
that incorporates the knowledge-enhanced representation: $u'_i = (1-\rho)u_i + \rho \mathcal{K}$.
The gradient with respect to parameters \(\theta_t\) is: $\nabla_{\theta_t} \mathcal{L} = \left( \frac{\partial \mathcal{L}}{\partial u_i} + (1-\rho) \frac{\partial \mathcal{L}}{\partial u'_i} \right) \frac{\partial u_i}{\partial \theta_t}$.
By performing a first-order Taylor expansion of the loss derivative around the ideal representation, we obtain:
\begin{equation}
\frac{\partial \mathcal{L}_{\text{actual}}}{\partial u'_i} - \frac{\partial \mathcal{L}_{\text{ideal}}}{\partial u'_i} \approx H_{\mathcal{L}}(u'_{\text{ideal}}) \cdot \rho \Delta \mathcal{K},
\end{equation}
where \(H_{\mathcal{L}}\) is the Hessian matrix. This proves that the gradient misalignment \(\Delta \nabla\) is approximately proportional to the knowledge deviation \(\Delta \mathcal{K}\). To bound this deviation, we leverage Pinsker's inequality to relate the KL divergence to the Total Variation Distance, yielding the relationship:
\begin{equation}
\label{eq:rdm_final_bound}
\|\Delta \mathcal{K}\|_2 \le C\sqrt{2 \cdot \text{RDM}(t, t_k)},
\end{equation}
where $C = \max_j \|z_j\|_2$ is a bounded constant. Equation \ref{eq:rdm_final_bound} establishes a formal link between RDM and the potential error margin for the gradients. Higher RDM directly widens this margin, causing training instability and representational noise. This theoretical foundation necessitates our dynamic refinement mechanism, which periodically resets the RDM to zero to ensure stable co-evolution between the model and its knowledge.

\begin{figure*}[htbp]
  \includegraphics[width=2\columnwidth]{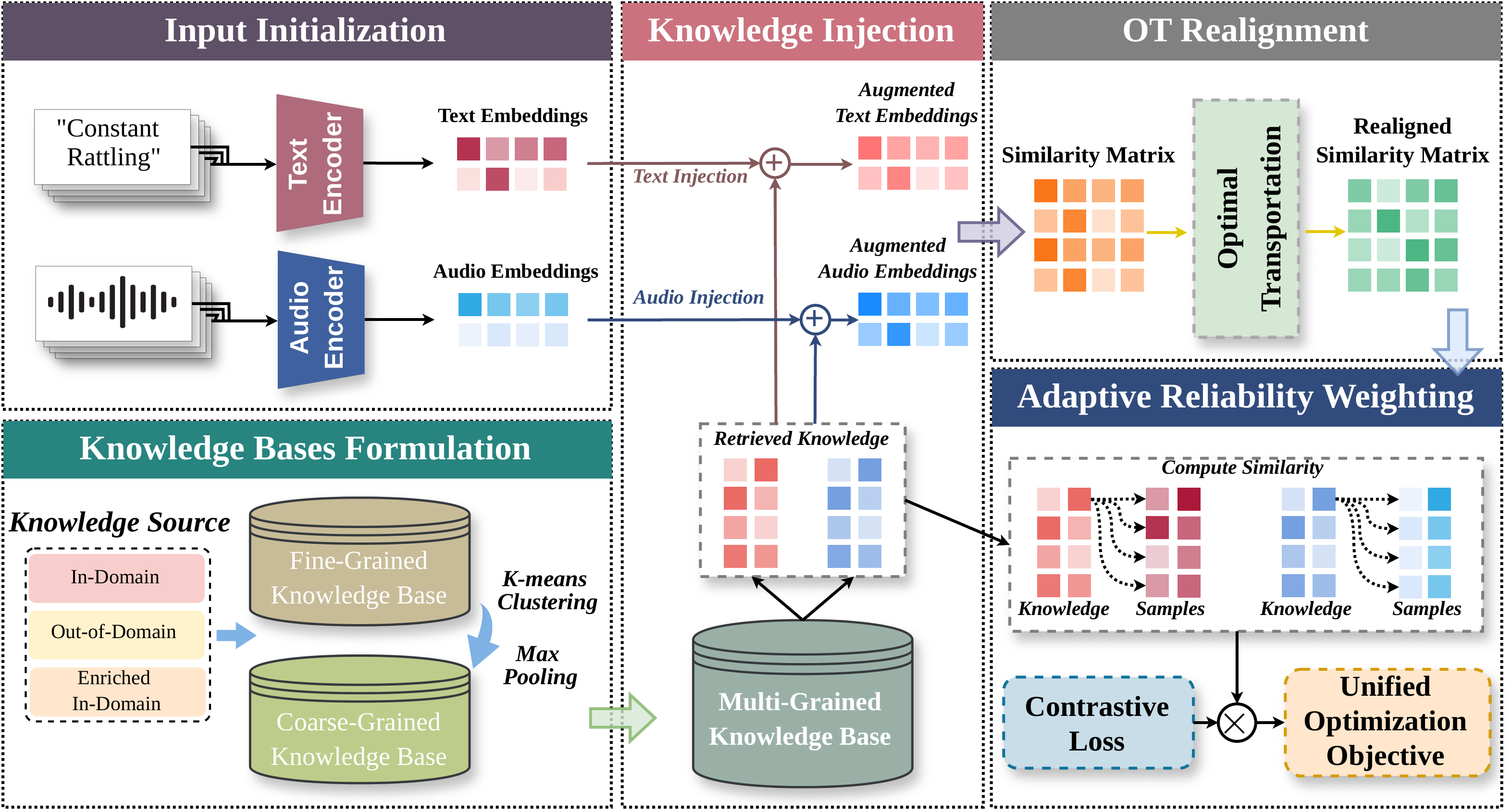}
  \caption{The proposed ASK framework. A multi-grained knowledge base is periodically updated to mitigate RDM. During training, knowledge is injected into samples, and a cross-modal reliability weight is computed. A final loss is optimized using both an OT-realigned similarity matrix and the reliability weight.}
  \label{fig:framework}
\end{figure*}

\section{The Adaptive Self-improving Knowledge Framework}
In this section, we elaborate on each component of our proposed framework ASK, whose architecture is shown in Figure~\ref{fig:framework}.

\subsection{Formulation of Knowledge Bases}
\label{sec:kb_formulation}

Our framework's first step is to construct multi-grained knowledge bases from a source dataset, $\mathcal{D}_k$. The choice of source is flexible; in our experiments, we explore three types to demonstrate versatility: 1) In-Domain $^+$ : the training set itself, 2) Out-of-Domain $^\dagger$: WavCaps \citep{mei2024wavcaps}, and 3) Enriched In-Domain $^*$: training set re-annotated by Gemini 2.5 \citep{comanici2025gemini}. From a chosen source, we construct two complementary bases.

\paragraph{\textbf{Fine-Grained Knowledge Base}.}
The fine-grained base, $K_f$, captures instance-level semantic details. It is formed by encoding all audio-text pairs in the source $\mathcal{D}_k = \{(a_j^k, t_j^k)\}_{j=1}^{N_k}$ using the current model encoders $f_\theta(\cdot)$ and $g_\phi(\cdot)$. The result is a collection of L2-normalized embedding pairs:
\begin{equation}
\begin{split}
    K_f &= \{ (u_j^k, v_j^k) \}_{j=1}^{N_k}, 
\end{split}
\end{equation}
where $u_j^k = f_\theta(a_j^k), v_j^k = g_\phi(t_j^k)$.

\paragraph{\textbf{Coarse-Grained Knowledge Base}.}
The coarse-grained base, $K_c$, provides a global semantic prior by storing a set of learned prototypes. These prototypes are generated by first partitioning the fine-grained embeddings via K-Means clustering into $N_c$ groups, and then distilling the salient features from each group. For the $m$-th audio cluster $\mathcal{C}_m^u$, which contains all member embeddings $\{u_j^k\}$, its prototype $c_m^u$ is computed via max-pooling:
\begin{equation}
    c_m^u = \text{MaxPooling}(\{ u_j^k \mid u_j^k \in \mathcal{C}_m^u \}).
\end{equation}
An identical procedure is applied to the text embeddings to yield text prototypes $\{c_m^v\}_{m=1}^{N_c}$. The final coarse-grained base is the set of these prototype pairs, $K_c = \{ (c_m^u, c_m^v) \}_{m=1}^{N_c}$.

\subsection{Multi-Grained Knowledge Injection}
\label{sec:knowledge_injection}

With the knowledge bases established, we perform two parallel injection processes to create distinct fine-grained and coarse-grained enhanced embeddings for each training sample. 

For the fine-grained injection, we first retrieve the Top-K nearest neighbors for a given embedding (e.g., audio $u_i$) from $K_f$, yielding the neighborhood set $\mathcal{N}_f(u_i)$. The retrieved embeddings are averaged to form a knowledge vector $\bar{u}_i^f$, which is then interpolated with the original embedding $u_i$:
\begin{equation}
\label{eq:fine_injection}
\begin{split}
u'_{i,f} &= (1-\rho) u_i + \rho \bar{u}_i^f, \\
\bar{u}_i^f &= \frac{\sum_{(u_j^k, v_j^k) \in \mathcal{N}_f(u_i)} u_j^k}{K},
\end{split}
\end{equation}
where $\rho$ is an interpolation hyperparameter. An identical, parallel process is performed using the coarse-grained base $K_c$ to produce the coarse-grained enhanced representation, $u'_{i,c}$. A symmetric procedure is applied to the text embedding $v_i$, ultimately yielding two distinct sets of enhanced embedding pairs for the final optimization: $(u'_{i,f}, v'_{i,f})$ and $(u'_{i,c}, v'_{i,c})$.

\paragraph{\textbf{Breaking the Gradient Locality Bottleneck}.}
This injection mechanism breaks the GLB (Sec.~\ref{sec:glb}) by creating a gradient pathway to out-of-batch knowledge. For any out-of-batch knowledge item $u_k^k$ retrieved by an in-batch sample $u_i$, its gradient is non-zero. Let $\mathcal{S}_k = \{i \in B \mid u_k^k \in \mathcal{N}_f(u_i)\}$ be the set of in-batch samples that retrieved $u_k^k$. The gradient of the loss $\mathcal{L}'_B$ w.r.t. $u_k^k$ is:
\begin{equation}
\label{eq:glb_break_grad}
\frac{\partial \mathcal{L}'_B}{\partial u_k^k} = \sum_{i \in \mathcal{S}_k} \frac{\partial \mathcal{L}'_B}{\partial u'_{i,f}} \frac{\partial u'_{i,f}}{\partial u_k^k}.
\end{equation}
From Eq.~\ref{eq:fine_injection}, the second partial derivative is a non-zero constant $\frac{\rho}{K}$. Given that one of its partial derivatives is non-zero, the total gradient is therefore non-zero. Consequently, the OBI, defined in Eq.~\ref{eq:obi}, becomes strictly positive. This quantitatively proves that our injection process breaks the GLB.

\subsection{Adaptive Reliability Weighting}
\label{sec:adaptive_weighting}

To mitigate the risk of injecting noisy knowledge from equally-weighted neighbors (Sec.~\ref{sec:knowledge_injection}), we introduce an adaptive weighting mechanism. This mechanism is based on the principle of cross-modal consistency: for a well-aligned audio-text pair $(u_i, v_i)$, the neighborhoods retrieved by $u_i$ and $v_i$ should themselves be semantically consistent. We quantify this consistency to compute a reliability score for each neighbor, which in turn modulates its contribution to the final objective.

\paragraph{\textbf{Fine-Grained Reliability Weighting}.}
For each pair $(u_i, v_i)$, we consider two fine-grained neighborhoods: the audio-retrieved audio set $\mathcal{U}_r = \{u_l^k\}_{l=1}^K$ and the text-retrieved audio–text set $\mathcal{N}_f(v_i)=\{(u_j^{k'}, v_j^k)\}_{j=1}^K$.
We first assign each neighbor in $\mathcal{N}_f(v_i)$ a consistency score $\bar{s}_j$, defined as its average similarity to the audio-retrieved neighborhood:
\begin{equation}
\bar{s}_j = \frac{1}{K} \sum_{l=1}^{K} (u_j^{k'})^\top u_l^k.
\end{equation}
These scores are subsequently normalized via a softmax function to yield the reliability weights $\mathbf{w}_f=\{w_j\}_{j=1}^K$:
\begin{equation}
\label{eq:weight_vector}
w_j = \frac{\exp(\bar{s}_j)}{\sum_{m=1}^{K}\exp(\bar{s}_m)}.
\end{equation}

The reliability-aware knowledge potential is then computed as the weighted similarity between $u_i$ and the audio components of $\mathcal{N}_f(v_i)$:
\begin{equation}
\Psi_{i,f}^{T \to A} = \sum_{j=1}^{K} w_j \cdot \exp(u_i^\top u_j^{k'}).
\end{equation}
A symmetric construction produces the text-side potential $\Psi_{i,f}^{A \to T}$, based on the audio-retrieved text neighborhood.

\paragraph{\textbf{Coarse-Grained Reliability Weighting}.}
An identical procedure is applied to the coarse-grained neighborhoods to produce the coarse-grained potentials, $\Psi_{i,c}^{T \to A}$ and $\Psi_{i,c}^{A \to T}$. These potentials represent the model's alignment with reliable, high-level semantic prototypes.

The resulting four reliability-aware potentials are core components that will be directly incorporated into our final optimization objective, as detailed in Section~\ref{sec:unified_objective}.

\subsection{Dynamic Knowledge Refinement}
\label{sec:dynamic_refinement}

As shown in Section~\ref{sec:rdm}, a static knowledge base leads to Representation Drift Mismatch (RDM), which induces increasing gradient misalignment during training. To mitigate this, we employ a dynamic refinement mechanism that periodically reconstructs the knowledge bases $K_f$ and $K_c$ using the current encoders. The update period $\mathcal{T}$ specifies the number of epochs between successive reconstructions.

This procedure directly controls the RDM. At each update step $t$, refinement sets the knowledge-base timestamp to $t_k = t$, making the ideal and actual neighborhood distributions identical, $P_{\text{ideal}} \equiv P_{\text{actual}}$. Thus, the RDM (Eq.~\ref{eq:rdm_definition}) is reset to its minimum value:

\begin{equation}
\text{RDM}(t, t) = \mathbb{E}[D_{KL}(P_{\text{ideal}} \,\|\, P_{\text{ideal}})] = 0.
\end{equation}
By periodically driving the RDM to zero, the mechanism also resets the upper bound on gradient deviation (Eq.~\ref{eq:rdm_final_bound}), ensuring stable optimization and enabling the knowledge base to co-evolve with the model.

\subsection{Unified Optimization Objective}
\label{sec:unified_objective}

The final optimization objective is constructed in two main stages. First, we compute NT-Xent losses on similarity matrices that have been realigned via Optimal Transport. Second, these losses are modulated by our reliability-aware knowledge potentials to form the final composite objective.

\paragraph{\textbf{Loss on OT-Realigned Similarities}.}
The process begins with the knowledge-enhanced embeddings from Section~\ref{sec:knowledge_injection}. For a mini-batch, we compute a fine-grained similarity matrix $\mathbf{S}_f$ and a coarse-grained one $\mathbf{S}_c$. Since the audio and text knowledge are retrieved independently, the distributions of their nearest neighbors within the batch may differ. To reconcile this potential discrepancy and find a globally optimal batch-level matching, we employ Optimal Transport (OT) \citep{cuturi2013sinkhorn} to learn an optimal transport plan $\mathbf{Q}^*$. This plan is then used to produce the realigned similarity matrices $\mathbf{S}^*_f$ and $\mathbf{S}^*_c$:
\begin{equation}
\label{eq:ot_align}
\mathbf{S}^{*}_f = \big((1-\beta)\mathbf{I}+\beta\,\mathbf{Q}^*\big)\,\mathbf{S}_f.
\end{equation}
An identical process is applied to $\mathbf{S}_c$. Based on these realigned matrices, we define two NT-Xent loss components. The text-to-audio loss, $\mathcal{L}_{T \to A}$, is the sum of the fine- and coarse-grained objectives:
\begin{equation}
\begin{aligned}
\mathcal{L}_{T \to A}
= &-\frac{1}{B}\sum_{i=1}^{B}
\log \frac{\exp((\mathbf{S}^{*}_{f})_{ii}/\tau)}
{\sum_{j=1}^{B}\exp((\mathbf{S}^{*}_{f})_{ij}/\tau)}\\
&-\frac{1}{B}\sum_{i=1}^{B}\log \frac{\exp((\mathbf{S}^{*}_{c})_{ii}/\tau)}{\sum_{j=1}^{B}\exp((\mathbf{S}^{*}_{c})_{ij}/\tau)}.\\
\end{aligned}
\end{equation}
The audio-to-text loss, $\mathcal{L}_{A \to T}$, is formulated symmetrically.

\paragraph{\textbf{Reliability-Aware Objective}.}
The OT-realigned losses above do not yet account for the cross-modal consistency of the retrieved knowledge. To incorporate this, we use the knowledge potentials computed in Section~\ref{sec:adaptive_weighting} as reliability modulators. We first define the reliability-aware terms, e.g., for the text-to-audio direction:
\begin{equation}
\begin{split}
\mathcal{F}_{f}^{T \to A} &= \frac{1}{|B|}\sum_{i=1}^{|B|} -\log \Psi_{i,f}^{T \to A}, \\
\mathcal{F}_{c}^{T \to A} &= \frac{1}{|B|}\sum_{i=1}^{|B|} -\log \Psi_{i,c}^{T \to A}.
\end{split}
\end{equation}
The final text-to-audio loss, $\mathcal{L}^*_{T \to A}$, is then the base OT-realigned loss, modulated by a weighted sum of these reliability terms:
\begin{equation}
\label{eq:final_modulated_loss}
\mathcal{L}^*_{T \to A} = (1 + \lambda_f \mathcal{F}_{f}^{T \to A} + \lambda_c \mathcal{F}_{c}^{T \to A}) \cdot \mathcal{L}_{T \to A},
\end{equation}
where $\lambda_f$ and $\lambda_c$ are hyperparameters. The final audio-to-text loss, $\mathcal{L}^*_{A \to T}$, is computed symmetrically. The overall loss for the ASK framework is the average of these two modulated objectives:
\begin{equation}
\mathcal{L}_{\text{ASK}} = \frac{1}{2}(\mathcal{L}^*_{T \to A} + \mathcal{L}^*_{A \to T}).
\end{equation}
This composite objective ensures the model learns from multi-grained knowledge that is both globally aligned at the batch level and weighted by its cross-modal reliability.

\subsection{Theoretical Analysis}
We provide a theoretical justification for ASK, framing the training as an alternating optimization to maximize the log-likelihood of observed audio-text pairs $x_i = (a_i, t_i)$.

\paragraph{\textbf{Probabilistic Formulation via ELBO}}
Our goal is to maximize the log-likelihood $\mathcal{L}(\theta) = \sum_i \log p(x_i; \theta)$. By introducing latent variables $z_i = (z_{i,f}, z_{i,c})$ representing the optimal knowledge, and an auxiliary distribution $Q(z_i)$, we apply Jensen's Inequality to derive the Evidence Lower Bound (ELBO), $\mathcal{F}(Q, \theta)$:
\begin{equation}
\begin{aligned}
\mathcal{L}(\theta) &= \sum_i \log \sum_{z_i} \frac{p(x_i, z_i; \theta)}{Q(z_i)} Q(z_i) \\
&\ge \sum_i \mathbb{E}_{Q(z_i)} [\log p(x_i, z_i; \theta)] + H(Q) \triangleq \mathcal{F}(Q, \theta),
\end{aligned}
\end{equation}
where $H(Q)$ is the entropy and maximizing $\mathcal{L}(\theta)$ is achieved by iteratively maximizing $\mathcal{F}(Q, \theta)$.

\paragraph{\textbf{Alternating Optimization}.}
Let $\theta_t$ be the parameters at iteration $t$. The process alternates between two stages:

\textit{Stage 1: Auxiliary Distribution Update.}
Fixing $\theta_t$, we approximate the optimal $Q_t(z_i)$ using the retrieved neighbors. We define the probability mass of $Q_t$ over a neighbor $z_j$ directly via our reliability weights (Eq.~\ref{eq:weight_vector}): $Q_{t,f}(z_j) := w_{j,f}(\theta_t)$ and $Q_{t,c}(z_j) := w_{j,c}(\theta_t)$.

\textit{Stage 2: Model Parameter Update.}
Fixing $Q_t$, we maximize the expectation $\mathbb{E}_{Q_t} [\log p(x_i, z_i; \theta)]$. We model the joint log-probability as the negative sum of the alignment loss and the reliability potential: $\log p(x_i, z_i; \theta) \propto -(\mathcal{L}_{OT}(\theta) + \log \Psi_{i}(\theta))$.
Substituting this into the ELBO, the maximization objective becomes minimizing the negative expectation:
\begin{equation}
\min_\theta \mathcal{L}_{m} \approx \sum_i \left( \mathbb{E}_{Q_{t,f}}[\mathcal{L}_{OT, f} + \log \Psi_{i,f}] + \mathbb{E}_{Q_{t,c}}[\mathcal{L}_{OT, c} + \log \Psi_{i,c}] \right).
\end{equation}
This objective $\mathcal{L}_{m}$ mathematically aligns with our modulated loss $\mathcal{L}^*$ (Eq.~\ref{eq:final_modulated_loss}), where the reliability term $\mathcal{F} = -\log \Psi$ acts as a regularizer.

\paragraph{\textbf{Convergence}.}
Since each step monotonically increases the ELBO $\mathcal{F}(Q, \theta)$, and $\mathcal{L}_{\text{ASK}}$ is bounded below, the Monotone Convergence Theorem guarantees that the sequence of loss values converges to a stationary point.

\section{Experiments}

\begin{table*}[t!]
    \centering
    \caption{Results for Audio-Text-Retrieval on AudioCaps and Clotho under the global interaction strategy. The symbols $^\dagger$, $^*$, and $^+$ denote the use of knowledge from WavCaps, the Gemini-annotated training set and the original training set respectively.}
    \label{tab:main_global_results_final}
    \renewcommand{\arraystretch}{1.2}
    \resizebox{2\columnwidth}{!}{
    \begin{tabular}{l ccc ccc ccc ccc}
        \toprule
        & \multicolumn{6}{c}{AudioCaps} & \multicolumn{6}{c}{Clotho} \\
        \cmidrule(lr){2-7} \cmidrule(lr){8-13}
        Method & \multicolumn{3}{c}{Audio-to-Text} & \multicolumn{3}{c}{Text-to-Audio} & \multicolumn{3}{c}{Audio-to-Text} & \multicolumn{3}{c}{Text-to-Audio} \\
        \cmidrule(lr){2-4} \cmidrule(lr){5-7} \cmidrule(lr){8-10} \cmidrule(lr){11-13}
        & R@1 & R@5 & R@10 & R@1 & R@5 & R@10 & R@1 & R@5 & R@10 & R@1 & R@5 & R@10 \\
        \midrule
        \rowcolor{gray!30!white} \multicolumn{13}{c}{Architecture: ResNet-38 + BERT} \\
        \midrule
        ML-ACT \cite{mei2022metric} & $36.3_{\pm 0.5}$ & $68.6_{\pm 0.3}$ & $81.5_{\pm 0.2}$ & $32.2_{\pm 0.4}$ & $68.2_{\pm 0.1}$ & $81.2_{\pm 0.2}$ & $16.3_{\pm 0.4}$ & $39.1_{\pm 0.3}$ & $51.5_{  \pm 0.6}$ & $14.2_{\pm 0.4}$ & $37.3_{ \pm 0.2}$ & $49.9_{  \pm 0.3}$ \\
        BLAT \cite{xu2023blat} & $38.2_{\pm 0.2}$ & $70.4_{\pm 0.3}$ & $82.1_{\pm 0.2}$ & $32.9_{\pm 0.3}$ & $68.9_{\pm 0.1}$ & $81.8_{\pm 0.2}$ & $16.8_{\pm 0.2}$ & $39.6_{\pm 0.3}$ & $52.1_{  \pm 0.3}$ & $14.1_{\pm 0.2}$ & $37.6_{ \pm 0.2}$ & $50.2_{  \pm 0.1}$ \\
        Auto-ACD \cite{sun2024autoacd} & $40.8_{\pm 0.2}$ & $71.3_{\pm 0.4}$ & $83.3_{\pm 0.2}$ & $33.2_{\pm 0.3}$ & $68.7_{\pm 0.2}$ & $82.1_{\pm 0.2}$ & $17.1_{\pm 0.2}$ & $39.3_{\pm 0.2}$ & $53.2_{  \pm 0.3}$ & $14.4_{\pm 0.2}$ & $37.5_{ \pm 0.2}$ & $50.1_{  \pm 0.2}$ \\
        \textbf{ASK}$^\dagger$  & $\textbf{42.3}_{\pm 0.3}$ & $73.3_{  \pm 0.8}$ & $84.2_{  \pm 0.6}$ & $34.6_{  \pm 0.5}$ & $69.6_{  \pm 0.4}$ & $82.9_{  \pm 0.9}$ & $17.3_{  \pm 0.3}$ & $40.2_{  \pm 0.6}$ & $\textbf{54.1}_{  \pm 0.2}$ & $14.8_{  \pm 0.4}$ & $38.1_{  \pm 0.7}$ & $50.7_{  \pm 0.6}$  \\
        \textbf{ASK}$^*$        & $39.5_{  \pm 0.3}$ & $73.2_{  \pm 0.4}$ & $85.3_{  \pm 0.6}$ & $34.2_{  \pm 0.6}$ & $69.1_{  \pm 0.7}$ & $81.9_{  \pm 0.3}$ & $\textbf{18.5}_{  \pm 0.2}$ & $40.1_{  \pm 0.4}$ & $53.6_{  \pm 0.6}$ & $14.7_{  \pm 0.5}$ & $38.3_{  \pm 0.9}$ & $50.1_{  \pm 0.3}$ \\
        \textbf{ASK}$^+$        & $42.0_{  \pm 0.2}$ & $\textbf{74.2}_{  \pm 0.5}$ & $\textbf{85.4}_{  \pm 0.6}$ & $\textbf{35.4}_{  \pm 0.3}$ & $\textbf{70.2}_{  \pm 0.3}$ & $\textbf{83.1}_{  \pm 0.7}$ & $17.5_{  \pm 0.2}$ & $\textbf{40.3}_{  \pm 0.8}$ & $\textbf{54.1}_{  \pm 0.6}$ & $\textbf{15.2}_{  \pm 0.3}$ & $\textbf{38.5}_{  \pm 0.6}$ & $\textbf{51.1}_{  \pm 0.4}$ \\
        \midrule
        \rowcolor{gray!30!white} \multicolumn{13}{c}{Architecture: CED-Base + SONAR-TE} \\
        \midrule
        ML-CLAP \cite{yan2024bridging} & $39.6_{  \pm 0.2}$ & $69.8_{  \pm 0.3}$ & $81.7_{  \pm 0.6}$ & $31.9_{  \pm 0.3}$ & $69.2_{  \pm 0.5}$ & $82.8_{  \pm 0.9}$ & $18.0_{  \pm 0.2}$ & $39.5_{  \pm 0.7}$ & $53.0_{  \pm 0.6}$ & $14.9_{  \pm 0.3}$ & $39.9_{  \pm 0.6}$ & $53.1_{  \pm 0.7}$ \\
        GLAP \cite{glap} & $41.2_{\pm 0.1}$ & $72.4_{\pm 0.3}$ & $83.7_{\pm 0.2}$ & $33.3_{\pm 0.3}$ & $68.9_{\pm 0.1}$ & $81.8_{\pm 0.4}$ & $18.4_{\pm 0.2}$ & $40.6_{\pm 0.3}$ & $54.1_{  \pm 0.3}$ & $15.1_{\pm 0.2}$ & $40.2_{ \pm 0.2}$ & $54.2_{  \pm 0.1}$ \\
        \textbf{ASK}$^\dagger$  & $\textbf{43.3}_{  \pm 0.3}$ & $73.7_{  \pm 0.6}$ & $84.4_{  \pm 0.8}$ & $34.8_{  \pm 0.2}$ & $70.6_{  \pm 0.5}$ & $84.0_{  \pm 0.4}$ & $19.0_{  \pm 0.1}$ & $41.5_{  \pm 0.6}$ & $56.5_{  \pm 0.7}$ & $\textbf{16.3}_{  \pm 0.2}$ & $40.3_{  \pm 0.6}$ & $\textbf{55.4}_{  \pm 0.7}$ \\
        \textbf{ASK}$^*$        & $41.9_{  \pm 0.3}$ & $\textbf{74.1}_{  \pm 0.4}$ & $\textbf{85.6}_{  \pm 0.6}$ & $\textbf{34.9}_{  \pm 0.2}$ & $\textbf{70.9}_{  \pm 0.5}$ & $\textbf{84.1}_{  \pm 0.7}$ & $18.5_{  \pm 0.2}$ & $41.6_{  \pm 0.6}$ & $56.9_{  \pm 0.7}$ & $16.0_{  \pm 0.1}$ & $40.6_{  \pm 0.5}$ & $55.1_{  \pm 0.8}$ \\
        \textbf{ASK}$^+$        & $40.9_{  \pm 0.2}$ & $71.6_{  \pm 0.6}$ & $84.3_{  \pm 0.3}$ & $33.7_{  \pm 0.2}$ & $70.3_{  \pm 0.5}$ & $83.5_{  \pm 0.6}$ & $\textbf{19.7}_{  \pm 0.1}$ & $\textbf{43.3}_{  \pm 0.5}$ & $\textbf{57.3}_{  \pm 0.7}$ & $16.0_{  \pm 0.2}$ & $\textbf{41.5}_{  \pm 0.6}$ & $\textbf{55.2}_{  \pm 0.7}$     \\
        \bottomrule
    \end{tabular}
    }
\end{table*}

\subsection{Experimental Setup}
\label{sec:experimental_setup}
\paragraph{\textbf{Datasets and Metrics}.}
We evaluate our method on two standard benchmarks: AudioCaps \citep{kim2019audiocaps} and Clotho \citep{drossos2020clotho}. Following prior work \citep{mei2022metric,xie2024gpa,yan2024bridging}, we report audio-to-text (A2T) and text-to-audio (T2A) retrieval performance using Recall at K (R@K, for K=1, 5, 10).

\paragraph{\textbf{Baselines}.}
To validate ASK's model-agnosticism, we evaluate it across two interaction paradigms. \textbf{1) Global Interaction:} We build upon the ML-ACT baseline \citep{mei2022metric} (ResNet-38 \citep{kong2020panns} + BERT \citep{devlin2019bert}), comparing against BLAT \citep{xu2023blat} and Auto-ACD \citep{sun2024autoacd}. We also adapt the English-only ML-CLAP setup \citep{yan2024bridging} (CED-Base \citep{dinkel2024ced} + SONAR-TE \citep{duquenne2023sonar}), comparing it with GLAP \citep{glap}. \textbf{2) Local Interaction:} We follow the setups of GPA \citep{xie2024gpa} and FLAM \citep{wu2025flam}, and adopt the same maximum number of tokens for the dataset.

\paragraph{\textbf{Implementation Details}.}
All models are trained with the Adam optimizer~\citep{kingma2014adam}. The ResNet-BERT architecture is trained for 50 epochs on AudioCaps (batch size 32) and Clotho (batch size 24), with an initial learning rate of $5 \times 10^{-5}$, which is decayed by a factor of 10 every 20 epochs. The CED-SONAR models are trained for 10 epochs with a decay step applied every 4 epochs. We use the Faiss library \citep{douze2025faiss} for efficient neighbor search. Unless specified otherwise, the hyperparameters for our ASK framework are set as follows: we retrieve $K=10$ neighbors, with a coarse-grained prototype set of size $N_c=512$. The knowledge injection ratio is $\rho=0.2$, and the OT-realignment factor is $\beta=0.2$. Empirical analyses confirm our framework's robustness across $K \in [5, 15]$ and $\rho \in [0.15, 0.25]$. The reliability modulation weights are $\lambda_f=0.2$ and $\lambda_c=0.3$. The knowledge base is dynamically refined every $\mathcal{T}=15$ epochs. All experiments were conducted on 2 NVIDIA A100 and 8 RTX 4090 GPUs.

\subsection{Main Results}

We evaluate the effectiveness of our proposed ASK framework by integrating it into various baseline models. The results are organized by the cross-modal interaction strategy.

\paragraph{\textbf{Global Interaction Strategy}.}
Table~\ref{tab:main_global_results_final} presents the results for models using a global, sentence-level interaction strategy. ASK consistently outperforms competitive methods across all datasets and architectures. On AudioCaps (ResNet-BERT), ASK surpasses the foundational ML-ACT by a remarkable 6.0\% (A2T) and 3.2\% (T2A) in absolute R@1. Crucially, it eclipses Auto-ACD, with further R@1 gains of up to 1.5\% (\textbf{ASK}$^\dagger$) and 2.2\% (\textbf{ASK}$^+$). These improvements validate our core mechanisms in successfully breaking the GLB and mitigating RDM. Demonstrating its model-agnosticism, ASK also significantly enhances the transformer-based CED-SONAR architecture. On the Clotho dataset, it outperforms the GLAP baseline, boosting A2T R@1 to 19.7\% (+1.3\%) and T2A R@1 to 16.3\% (+1.2\%). Finally, the varying optimal variants across setups highlight ASK's flexibility in leveraging diverse knowledge sources.

\begin{figure}[htbp]
\centering
  \includegraphics[width=\linewidth]{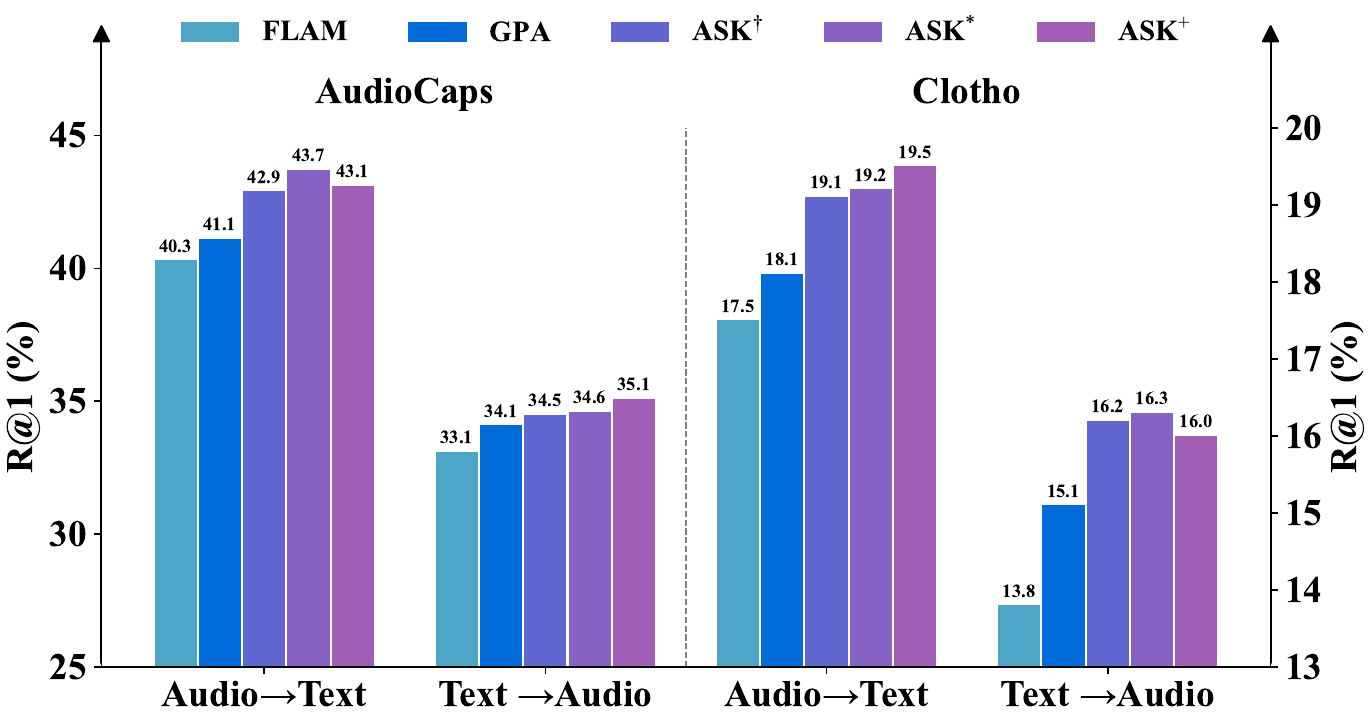}
  \caption{Results for Audio-Text Retrieval on AudioCaps and Clotho under the local interaction strategy. The symbols $^+$, $^\dagger$, and $^*$ denote different knowledge sources in Section~\ref{sec:kb_formulation}.}
  \label{fig:local}
\end{figure}

\paragraph{\textbf{Local Interaction Strategy}.}

We also validate ASK with local, token-level interaction strategies \citep{xie2024gpa, wu2025flam}. 
Figure~\ref{fig:local} presents the retrieval results for our local, token-level interaction baselines. The results demonstrate consistent and significant gains across both retrieval directions, outperforming both the FLAM and GPA baselines. On AudioCaps, ASK achieves substantial improvements over the strongest baseline (GPA), with \textbf{ASK}$^*$ enhancing the Audio-to-Text R@1 by 2.6\% and \textbf{ASK}$^+$ boosting the Text-to-Audio R@1 by 1.0\%. On Clotho, \textbf{ASK}$^+$ delivers the top Audio-to-Text R@1 performance, while \textbf{ASK}$^*$ yields the best Text-to-Audio R@1. Notably, our variants exceed FLAM by even larger margins across all metrics. These symmetric improvements confirm the universal benefit of our framework.

\begin{figure}[htbp]
\centering
  \includegraphics[width=\linewidth]{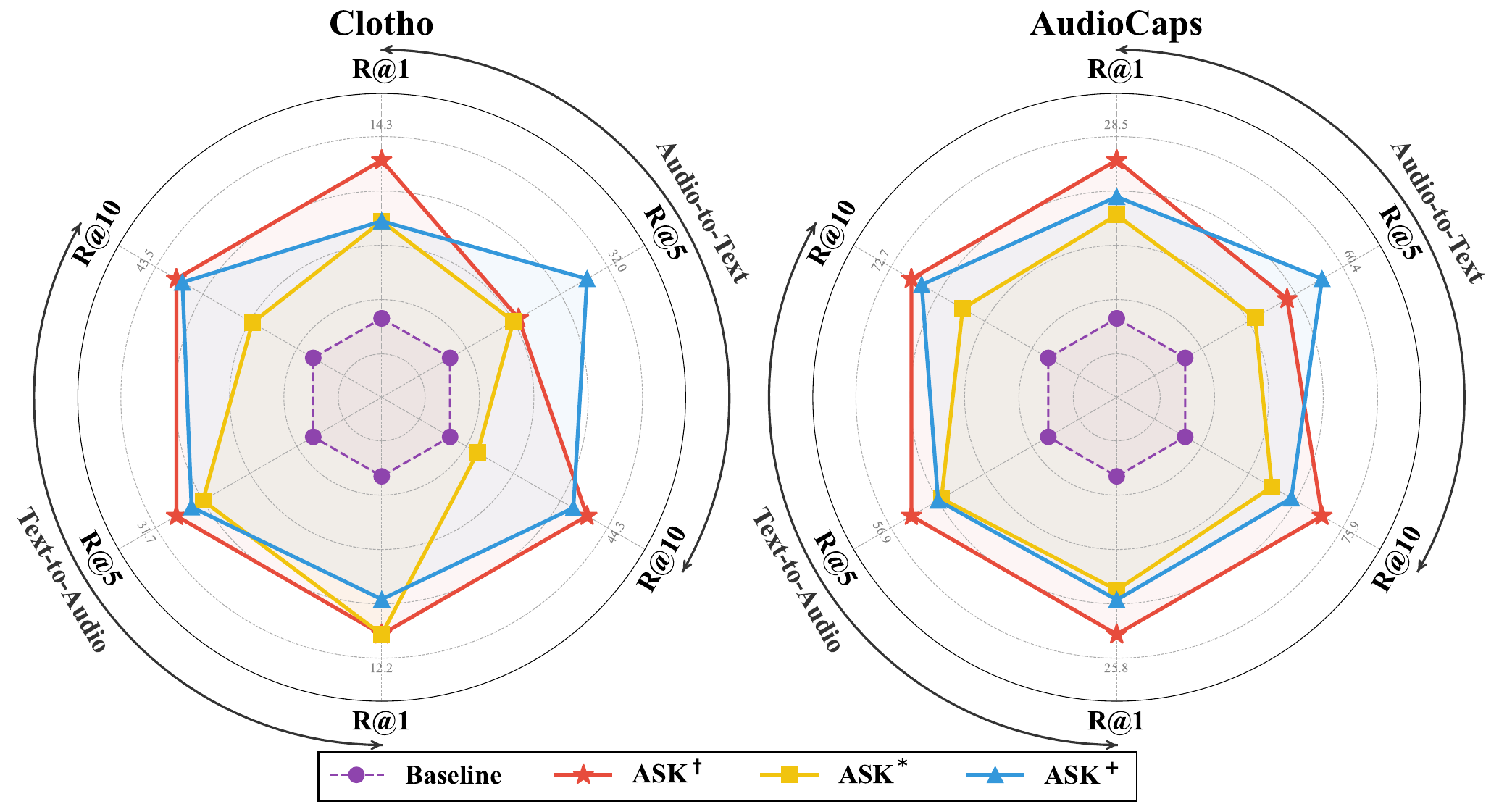}
  \caption{Zero-shot performance on AudioCaps and Clotho. $^\dagger$, $^*$, and $^+$ denote different knowledge sources in Section~\ref{sec:kb_formulation}.}
  \label{fig:zeroshot}
\end{figure}

\paragraph{\textbf{Zero-Shot Generalization}.} 
To strictly evaluate robustness, we conduct bidirectional zero-shot experiments: training on AudioCaps and testing on Clotho, and vice versa. As illustrated in Figure~\ref{fig:zeroshot}, the ASK framework consistently expands the retrieval performance envelope compared to the baseline across all metrics in both transfer directions. 
Specifically, when transferring from AudioCaps to Clotho (left), leveraging the diverse WavCaps knowledge source (ASK$^\dagger$) yields a notable \textbf{1.3\%} absolute gain in A2T R@1. Similarly, in the Clotho-to-AudioCaps transfer (right), ASK variants maintain superior performance, with ASK$^\dagger$ achieving a significant improvement over the baseline (e.g., \textbf{+2.6\%} in A2T R@1). These results confirm that our multi-grained knowledge injection effectively mitigates domain shifts and prevents overfitting to source-domain specifics.

\begin{table*}[t!] 
    \centering
    \caption{Ablation experiments on AudioCaps dataset using the ResNet-38 + BERT architecture. $^+$ denotes the utilization of knowledge derived from AudioCaps training set.} 
    \label{tab:ablation_study}
    \resizebox{2\columnwidth}{!}{
    
    \begin{tabular}{ll ccc ccc}
        \toprule
        & & \multicolumn{3}{c}{A2T} & \multicolumn{3}{c}{T2A} \\
        \cmidrule(lr){3-5} \cmidrule(lr){6-8}
        G. & Method & R@1 & R@5 & R@10 & R@1 & R@5 & R@10 \\
        \midrule
        & w/o ASK  & $36.3_{  \pm 0.5}$ & $68.6_{  \pm 0.3}$ & $81.5_{  \pm 0.2}$ & $32.2_{  \pm 0.4}$ & $68.2_{  \pm 0.1}$ & $81.2_{  \pm 0.2}$ \\
        \midrule
        1 & w/o Fine-grained Knowledge Base  & $37.7_{  \pm 0.2}$ & $70.4_{  \pm 0.4}$ & $81.8_{  \pm 0.7}$ & $31.9_{  \pm 0.2}$ & $67.3_{  \pm 0.6}$ & $81.0_{  \pm 0.7}$  \\
        & w/o Coarse-grained Knowledge Base  & $37.4_{  \pm 0.1}$ & $67.6_{  \pm 0.5}$ & $81.3_{  \pm 0.7}$ & $31.2_{  \pm 0.3}$ & $66.6_{  \pm 0.4}$ & $81.0_{  \pm 0.6}$   \\
        \midrule
        2 & w/o the Knowledge Injection Step & $39.1_{  \pm 0.3}$ & $72.7_{  \pm 0.6}$ & $84.1_{  \pm 0.7}$ & $34.5_{  \pm 0.3}$ & $69.1_{  \pm 0.6}$ & $82.6_{  \pm 0.7}$  \\
        & w/o OT Alignment Correction & $41.1_{  \pm 0.3}$ & $73.4_{  \pm 0.5}$ & $85.2_{  \pm 0.6}$ & $34.2_{  \pm 0.2}$ & $69.4_{  \pm 0.4}$ & $82.8_{  \pm 0.6}$ \\
        \midrule
        3 & w/o Adaptive Reliability Weighting & $39.3_{  \pm 0.2}$ & $72.2_{  \pm 0.4}$ & $83.6_{  \pm 0.6}$ & $33.9_{  \pm 0.3}$ & $68.9_{  \pm 0.5}$ & $81.6_{  \pm 0.7}$ \\
        \midrule
        4 & w/o the Dynamic Knowledge Refinement & $39.2_{  \pm 0.3}$ & $71.0_{  \pm 0.6}$ & $83.8_{  \pm 0.5}$ & $34.1_{  \pm 0.2}$ & $68.7_{  \pm 0.6}$ & $81.5_{  \pm 0.5}$ \\
        \midrule
         & Our Full \textbf{ASK}$^+$ & $\textbf{42.0}_{  \pm 0.2}$ & $\textbf{74.2}_{  \pm 0.5}$ & $\textbf{85.4}_{  \pm 0.6}$ & $\textbf{35.4}_{  \pm 0.3}$ & $\textbf{70.2}_{  \pm 0.3}$ & $\textbf{83.1}_{  \pm 0.7}$ \\
        \bottomrule
    \end{tabular}
    }
\end{table*}

\subsection{Ablation Study and Analysis}
\label{sec:ablation}

To validate the contribution of each component within our ASK framework, we conduct a series of ablation studies on the AudioCaps dataset using the ResNet-BERT architecture and an in-domain knowledge source. The results are presented in Table~\ref{tab:ablation_study}.

\paragraph{\textbf{Impact of Multi-Grained Knowledge Bases}.}
We first analyze the necessity of our multi-grained design. Removing the fine-grained knowledge base results in a substantial performance drop of 4.3\% absolute in A2T R@1, confirming the critical role of instance-level details for precise retrieval. Similarly, removing the coarse-grained base leads to a 4.6\% drop in A2T R@1, which underscores the importance of the global semantic prior provided by the prototypes. The model, which leverages both, significantly outperforms either single-granularity variant, demonstrating that the fine- and coarse-grained knowledge sources are complementary.

\begin{figure}[htbp]
\centering
  \includegraphics[width=\linewidth]{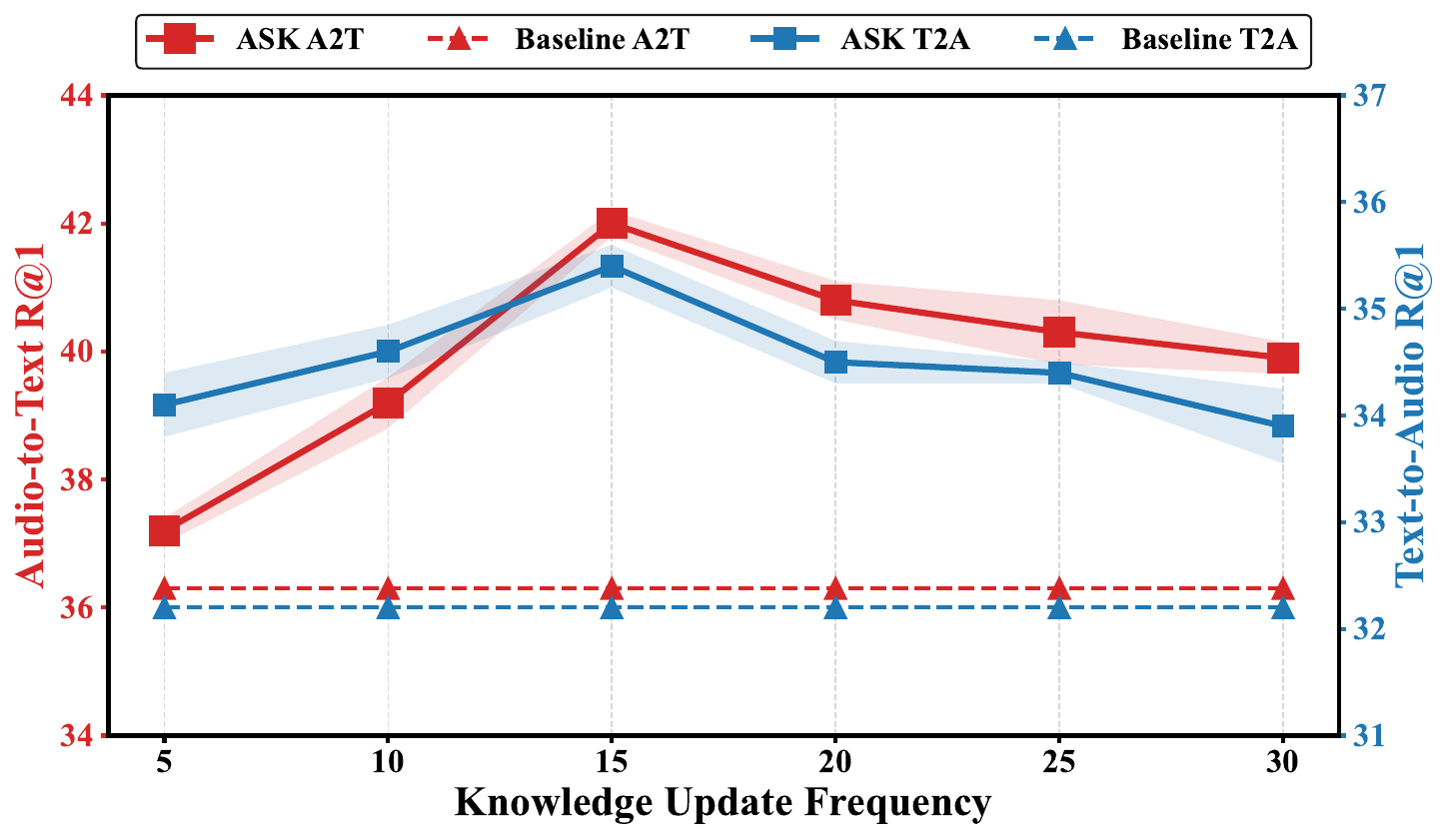}
  \caption{ Effect of the frequency $\mathcal{T}$ of Knowledge Update. Ablation experiment on \textbf{ASK}$^+$ under the global interaction strategy.
  }
  \label{fig:experiments}
\end{figure}

\paragraph{\textbf{Impact of Core ASK Mechanisms}.}
We then ablate the core mechanisms of ASK. 
\textbf{1) Knowledge Injection:} Disabling the knowledge injection step causes a notable drop of 2.9\% in A2T R@1. This empirically validates that creating gradient pathways to out-of-batch data is the primary driver for breaking the GLB and enhancing representations.
\textbf{2) Reliability Weighting:} Ablating our adaptive reliability weighting mechanism results in a significant 2.7\% drop in A2T R@1 and a 1.5\% drop in T2A R@1. This provides strong evidence that not all retrieved knowledge is equally beneficial, and that modulating the loss based on cross-modal consistency is crucial for mitigating the impact of noises and achieving robust performance.

\paragraph{\textbf{Impact of Dynamic Knowledge Refinement}.}
We evaluate the effect of the knowledge-base update period $\mathcal{T}$ on mitigating RDM. As shown in Table~\ref{tab:ablation_study}, disabling dynamic refinement leads to a 2.8\% drop in A2T R\@1, empirically validating our theoretical claim in Section~\ref{sec:rdm} that unchecked RDM introduces stale and misaligned knowledge.

Figure~\ref{fig:experiments} shows that performance improves as the update frequency increases, reaching an optimum at $T=15$ epochs, which surpasses both the static knowledge base and the baseline. However, overly frequent updates degrade performance, indicating a trade-off: while frequent updates curtail RDM, they can also destabilize the knowledge representation before the model fully adapts. These findings highlight the necessity of a co-evolving knowledge base and careful tuning of the update frequency.

\begin{table}[htbp]
\centering
\caption{Qualitative comparison of audio-text retrieval. GT represents Ground Truth.}
\label{tab:case_study}
\resizebox{\linewidth}{!}{%
\begin{tabular}{p{0.3\linewidth} p{0.3\linewidth} p{0.35\linewidth}}
\toprule
\multicolumn{1}{c}{\textbf{Audio Query}} & \textbf{ML-ACT \cite{mei2022metric}} & \textbf{ASK (Ours)} \\
\midrule
\rowcolor{gray!30!white} \multicolumn{3}{c}{\textit{Scenario 1: Semantic Ambiguity}} \\
\midrule
\textbf{GT:} \textit{``Thunder roars in the distance as rain falls''} \par \textit{(Confusing Texture)} 
& 
\textbf{Top-1:} \textit{``Food sizzling in a pan''} \xmark \par
(Audio Ambiguity) \par
\textbf{GT Rank: 21} 
& 
\textbf{Top-1:} \textit{``Thunder roars in the distance as rain falls''} \cmark \par
\textbf{GT Rank: 1} \\ 
\midrule
\rowcolor{gray!30!white} \multicolumn{3}{c}{\textit{Scenario 2: Long-tail Concept}} \\
\midrule
\textbf{GT:} \textit{``Church bells ringing''} \par \textit{(Rare Event)} 
& 
\textbf{Top-1:} \textit{``Train horn blowing''} \xmark \par
(Tonal Similarity) \par
\textbf{GT Rank: 13} 
& 
\textbf{Top-1:} \textit{``Church bells ringing''} \cmark \par
\textbf{GT Rank: 1} \\ 
\bottomrule
\end{tabular}%
}
\end{table}

\paragraph{\textbf{Case Study}.}
Table~\ref{tab:case_study} qualitatively illustrates how the ASK framework overcomes the Gradient Locality Bottleneck (GLB). In Scenario 1, the baseline model fails to resolve semantic ambiguity between acoustically similar textures like \textit{thunder} and \textit{sizzling food}, whereas ASK distinguishes these fine-grained details to achieve Rank 1. Likewise, for the long-tail \textit{Church bells} concept, ASK avoids the baseline's tendency to default to common tonal relatives like \textit{train horns}. These results demonstrate that leveraging diverse out-of-batch knowledge during training allows ASK to learn a discriminative embedding space that generalizes to rare and ambiguous events without requiring explicit retrieval during inference.

\section{Conclusion}
\label{sec:conclusion}

In this paper, we identified and formalized two fundamental challenges in knowledge-enhanced Audio-Text Retrieval: the Gradient Locality Bottleneck, which confines standard contrastive learning to mini-batches, and the consequent Representation-Drift Mismatch, which arises from using static knowledge bases with evolving models. To address this dual challenge, we proposed the Adaptive Self-improving Knowledge framework. ASK is a model-agnostic, plug-and-play solution that breaks the GLB via multi-grained knowledge injection, mitigates RDM through dynamic knowledge refinement, and ensures reliability with a novel adaptive weighting scheme. Extensive experiments demonstrate that ASK consistently and significantly improves performance across diverse architectures and datasets, achieving new state-of-the-art results.

\normalsize

\bibliography{main}

@inproceedings{mei2022metric,
  title={On Metric Learning for Audio-Text Cross-Modal Retrieval},
  author={Mei, Xinhao and Liu, Xubo and Sun, Jianyuan and Plumbley, Mark and Wang, Wenwu},
  booktitle={Proc. Interspeech 2022},
  pages={4142--4146},
  year={2022}
}

@inproceedings{devlin2019bert,
  title={Bert: Pre-training of deep bidirectional transformers for language understanding},
  author={Devlin, Jacob and Chang, Ming-Wei and Lee, Kenton and Toutanova, Kristina},
  booktitle={Proceedings of the 2019 conference of the North American chapter of the association for computational linguistics: human language technologies, volume 1 (long and short papers)},
  pages={4171--4186},
  year={2019}
}

@article{kong2020panns,
  title={Panns: Large-scale pretrained audio neural networks for audio pattern recognition},
  author={Kong, Qiuqiang and Cao, Yin and Iqbal, Turab and Wang, Yuxuan and Wang, Wenwu and Plumbley, Mark D},
  journal={IEEE/ACM Transactions on Audio, Speech, and Language Processing},
  volume={28},
  pages={2880--2894},
  year={2020},
  publisher={IEEE}
}

@inproceedings{xie2024gpa,
  title={GPA: Global and Prototype Alignment for Audio-Text Retrieval},
  author={Xie, Yuxin and Zhu, Zhihong and Zhuang, Xianwei and Liang, Liming and Wang, Zhichang and Zou, Yuexian},
  booktitle={Proc. Interspeech 2024},
  pages={5078--5082},
  year={2024}
}

@article{khandelwal2019generalization,
  title={Generalization through memorization: Nearest neighbor language models},
  author={Khandelwal, Urvashi and Levy, Omer and Jurafsky, Dan and Zettlemoyer, Luke and Lewis, Mike},
  journal={arXiv preprint arXiv:1911.00172},
  year={2019}
}

@article{cuturi2013sinkhorn,
  title={Sinkhorn distances: Lightspeed computation of optimal transport},
  author={Cuturi, Marco},
  journal={Advances in neural information processing systems},
  volume={26},
  year={2013}
}

@inproceedings{su2017order,
  title={Order-preserving wasserstein distance for sequence matching},
  author={Su, Bing and Hua, Gang},
  booktitle={Proceedings of the IEEE conference on computer vision and pattern recognition},
  pages={1049--1057},
  year={2017}
}

@inproceedings{yan2024bridging,
  title={Bridging Language Gaps in Audio-Text Retrieval},
  author={Yan, Zhiyong and Dinkel, Heinrich and Wang, Yongqing and Liu, Jizhong and Zhang, Junbo and Wang, Yujun and Wang, Bin},
  booktitle={Proc. Interspeech 2024},
  pages={1675--1679},
  year={2024}
}

@inproceedings{dinkel2024ced,
  title={CED: Consistent ensemble distillation for audio tagging},
  author={Dinkel, Heinrich and Wang, Yongqing and Yan, Zhiyong and Zhang, Junbo and Wang, Yujun},
  booktitle={ICASSP 2024-2024 IEEE International Conference on Acoustics, Speech and Signal Processing (ICASSP)},
  pages={291--295},
  year={2024},
  organization={IEEE}
}

@article{duquenne2023sonar,
  title={SONAR: sentence-level multimodal and language-agnostic representations},
  author={Duquenne, Paul-Ambroise and Schwenk, Holger and Sagot, Beno{\^\i}t},
  journal={arXiv preprint arXiv:2308.11466},
  year={2023}
}

@article{kingma2014adam,
  title={Adam: A method for stochastic optimization},
  author={Kingma, Diederik P and Ba, Jimmy},
  journal={arXiv preprint arXiv:1412.6980},
  year={2014}
}

@inproceedings{chen2020simple,
  title={A simple framework for contrastive learning of visual representations},
  author={Chen, Ting and Kornblith, Simon and Norouzi, Mohammad and Hinton, Geoffrey},
  booktitle={International conference on machine learning},
  pages={1597--1607},
  year={2020},
  organization={PmLR}
}

@article{kullback1951information,
  title={On information and sufficiency},
  author={Kullback, Solomon and Leibler, Richard A},
  journal={The annals of mathematical statistics},
  volume={22},
  number={1},
  pages={79--86},
  year={1951},
  publisher={JSTOR}
}

@article{mei2024wavcaps,
  title={Wavcaps: A chatgpt-assisted weakly-labelled audio captioning dataset for audio-language multimodal research},
  author={Mei, Xinhao and Meng, Chutong and Liu, Haohe and Kong, Qiuqiang and Ko, Tom and Zhao, Chengqi and Plumbley, Mark D and Zou, Yuexian and Wang, Wenwu},
  journal={IEEE/ACM Transactions on Audio, Speech, and Language Processing},
  volume={32},
  pages={3339--3354},
  year={2024},
  publisher={IEEE}
}

@article{comanici2025gemini,
  title={Gemini 2.5: Pushing the frontier with advanced reasoning, multimodality, long context, and next generation agentic capabilities},
  author={Comanici, Gheorghe and Bieber, Eric and Schaekermann, Mike and Pasupat, Ice and Sachdeva, Noveen and Dhillon, Inderjit and Blistein, Marcel and Ram, Ori and Zhang, Dan and Rosen, Evan and others},
  journal={arXiv preprint arXiv:2507.06261},
  year={2025}
}

@inproceedings{kim2019audiocaps,
  title={Audiocaps: Generating captions for audios in the wild},
  author={Kim, Chris Dongjoo and Kim, Byeongchang and Lee, Hyunmin and Kim, Gunhee},
  booktitle={Proceedings of the 2019 Conference of the North American Chapter of the Association for Computational Linguistics: Human Language Technologies, Volume 1 (Long and Short Papers)},
  pages={119--132},
  year={2019}
}

@inproceedings{drossos2020clotho,
  title={Clotho: An audio captioning dataset},
  author={Drossos, Konstantinos and Lipping, Samuel and Virtanen, Tuomas},
  booktitle={ICASSP 2020-2020 IEEE International Conference on Acoustics, Speech and Signal Processing (ICASSP)},
  pages={736--740},
  year={2020},
  organization={IEEE}
}

@article{douze2025faiss,
  title={The faiss library},
  author={Douze, Matthijs and Guzhva, Alexandr and Deng, Chengqi and Johnson, Jeff and Szilvasy, Gergely and Mazar{\'e}, Pierre-Emmanuel and Lomeli, Maria and Hosseini, Lucas and J{\'e}gou, Herv{\'e}},
  journal={IEEE Transactions on Big Data},
  year={2025},
  publisher={IEEE}
}

@inproceedings{guu2020retrieval,
  title={Retrieval augmented language model pre-training},
  author={Guu, Kelvin and Lee, Kenton and Tung, Zora and Pasupat, Panupong and Chang, Mingwei},
  booktitle={International conference on machine learning},
  pages={3929--3938},
  year={2020},
  organization={PMLR}
}

@article{Word2Vec,
  title={Distributed representations of words and phrases and their compositionality},
  author={Mikolov, Tomas and Sutskever, Ilya and Chen, Kai and Corrado, Greg S and Dean, Jeff},
  journal={Advances in neural information processing systems},
  volume={26},
  year={2013}
}

@article{MFCCs,
  title={Feature extraction with mel scale separation method on noise audio recordings},
  author={Roy Rudolf Huizen and Florentina Tatrin Kurniati},
  journal={arXiv preprint arXiv:2112.14930},
  year={2021},
 }

@article{ast,
  title={AST: Audio Spectrogram Transformer},
  author={Gong, Yuan and Chung, Yu-An and Glass, James},
  year={2021},
  journal={arXiv preprint arXiv:2104.01778}
}

@inproceedings{clap,
  title={Clap: Contrastive language-audio pre-training model for multi-modal sentiment analysis},
  author={Zhao, Tianqi and Kong, Ming and Liang, Tian and Zhu, Qiang and Kuang, Kun and Wu, Fei},
  booktitle={Proceedings of the 2023 ACM international conference on multimedia retrieval},
  pages={622--626},
  year={2023}
}

@inproceedings{audioclip,
  title={Audioclip: Extending clip to image, text and audio},
  author={Guzhov, Andrey and Raue, Federico and Hees, J{\"o}rn and Dengel, Andreas},
  booktitle={ICASSP 2022-2022 IEEE International Conference on Acoustics, Speech and Signal Processing (ICASSP)},
  pages={976--980},
  year={2022},
  organization={IEEE}
}

@inproceedings{wav2clip,
  title={Wav2clip: Learning robust audio representations from clip},
  author={Wu, Ho-Hsiang and Seetharaman, Prem and Kumar, Kundan and Bello, Juan Pablo},
  booktitle={ICASSP 2022-2022 IEEE International Conference on Acoustics, Speech and Signal Processing (ICASSP)},
  pages={4563--4567},
  year={2022},
  organization={IEEE}
}

@article{glap,
  title={GLAP: General Contrastive Audio–Text Pretraining Across Domains and Languages},
  author={Dinkel, Heinrich and others},
  year={2025},
  journal={arXiv preprint arXiv:2506.11350}
}

@inproceedings{clip,
  title={Learning transferable visual models from natural language supervision},
  author={Radford, Alec and Kim, Jong Wook and Hallacy, Chris and Ramesh, Aditya and Goh, Gabriel and Agarwal, Sandhini and Sastry, Girish and Askell, Amanda and Mishkin, Pamela and Clark, Jack and others},
  booktitle={International conference on machine learning},
  pages={8748--8763},
  year={2021},
  organization={PmLR}
}

@inproceedings{scan,
  title={Stacked cross attention for image-text matching},
  author={Lee, Kuang-Huei and Chen, Xi and Hua, Gang and Hu, Houdong and He, Xiaodong},
  booktitle={Proceedings of the European conference on computer vision (ECCV)},
  pages={201--216},
  year={2018}
}

@article{lu2019vilbert,
  title={Vilbert: Pretraining task-agnostic visiolinguistic representations for vision-and-language tasks},
  author={Lu, Jiasen and Batra, Dhruv and Parikh, Devi and Lee, Stefan},
  journal={Advances in neural information processing systems},
  volume={32},
  year={2019}
}

@inproceedings{yin2025atri,
  title={Atri: Mitigating multilingual audio text retrieval inconsistencies by reducing data distribution errors},
  author={Yin, Yuguo and Xie, Yuxin and Yang, Wenyuan and Yang, Dongchao and Ru, Jinghan and Zhuang, Xianwei and Liang, Liming and Zou, Yuexian},
  booktitle={Proceedings of the 63rd Annual Meeting of the Association for Computational Linguistics (Volume 1: Long Papers)},
  pages={5491--5504},
  year={2025}
}

@inproceedings{he2020momentum,
  title={Momentum contrast for unsupervised visual representation learning},
  author={He, Kaiming and Fan, Haoqi and Wu, Yuxin and Xie, Saining and Girshick, Ross},
  booktitle={Proceedings of the IEEE/CVF conference on computer vision and pattern recognition},
  pages={9729--9738},
  year={2020}
}

@inproceedings{xiong2020approximate,
  title={Approximate Nearest Neighbor Negative Contrastive Learning for Dense Text Retrieval},
  author={Xiong, Lee and Xiong, Chenyan and Li, Ye and Tang, Kwok-Fung and Liu, Jialin and Bennett, Paul N and Ahmed, Junaid and Overwijk, Arnold},
  booktitle={International Conference on Learning Representations}
}

@inproceedings{dwibedi2021little,
  title={With a little help from my friends: Nearest-neighbor contrastive learning of visual representations},
  author={Dwibedi, Debidatta and Aytar, Yusuf and Tompson, Jonathan and Sermanet, Pierre and Zisserman, Andrew},
  booktitle={Proceedings of the IEEE/CVF international conference on computer vision},
  pages={9588--9597},
  year={2021}
}

@inproceedings{qu2021rocketqa,
  title={RocketQA: An optimized training approach to dense passage retrieval for open-domain question answering},
  author={Qu, Yingqi and Ding, Yuchen and Liu, Jing and Liu, Kai and Ren, Ruiyang and Zhao, Wayne Xin and Dong, Daxiang and Wu, Hua and Wang, Haifeng},
  booktitle={Proceedings of the 2021 conference of the North American chapter of the association for computational linguistics: human language technologies},
  pages={5835--5847},
  year={2021}
}

@inproceedings{wu2025flam,
  title={FLAM: Frame-Wise Language-Audio Modeling},
  author={Wu, Yusong and Tsirigotis, Christos and Chen, Ke and Huang, Cheng-Zhi Anna and Courville, Aaron and Nieto, Oriol and Seetharaman, Prem and Salamon, Justin},
  booktitle={Forty-second International Conference on Machine Learning}
}

@inproceedings{sun2024autoacd,
  title={Auto-acd: A large-scale dataset for audio-language representation learning},
  author={Sun, Luoyi and Xu, Xuenan and Wu, Mengyue and Xie, Weidi},
  booktitle={Proceedings of the 32nd ACM International Conference on Multimedia},
  pages={5025--5034},
  year={2024}
}

@inproceedings{xu2023blat,
  title={Blat: Bootstrapping language-audio pre-training based on audioset tag-guided synthetic data},
  author={Xu, Xuenan and Zhang, Zhiling and Zhou, Zelin and Zhang, Pingyue and Xie, Zeyu and Wu, Mengyue and Zhu, Kenny Q},
  booktitle={Proceedings of the 31st ACM International Conference on Multimedia},
  pages={2756--2764},
  year={2023}
}

\appendix

\section{Derivation and Visualization of RDM's Impact}
\label{app:rdm_derivation}

\begin{figure}[h!]
\centering
  \includegraphics[width=\columnwidth]{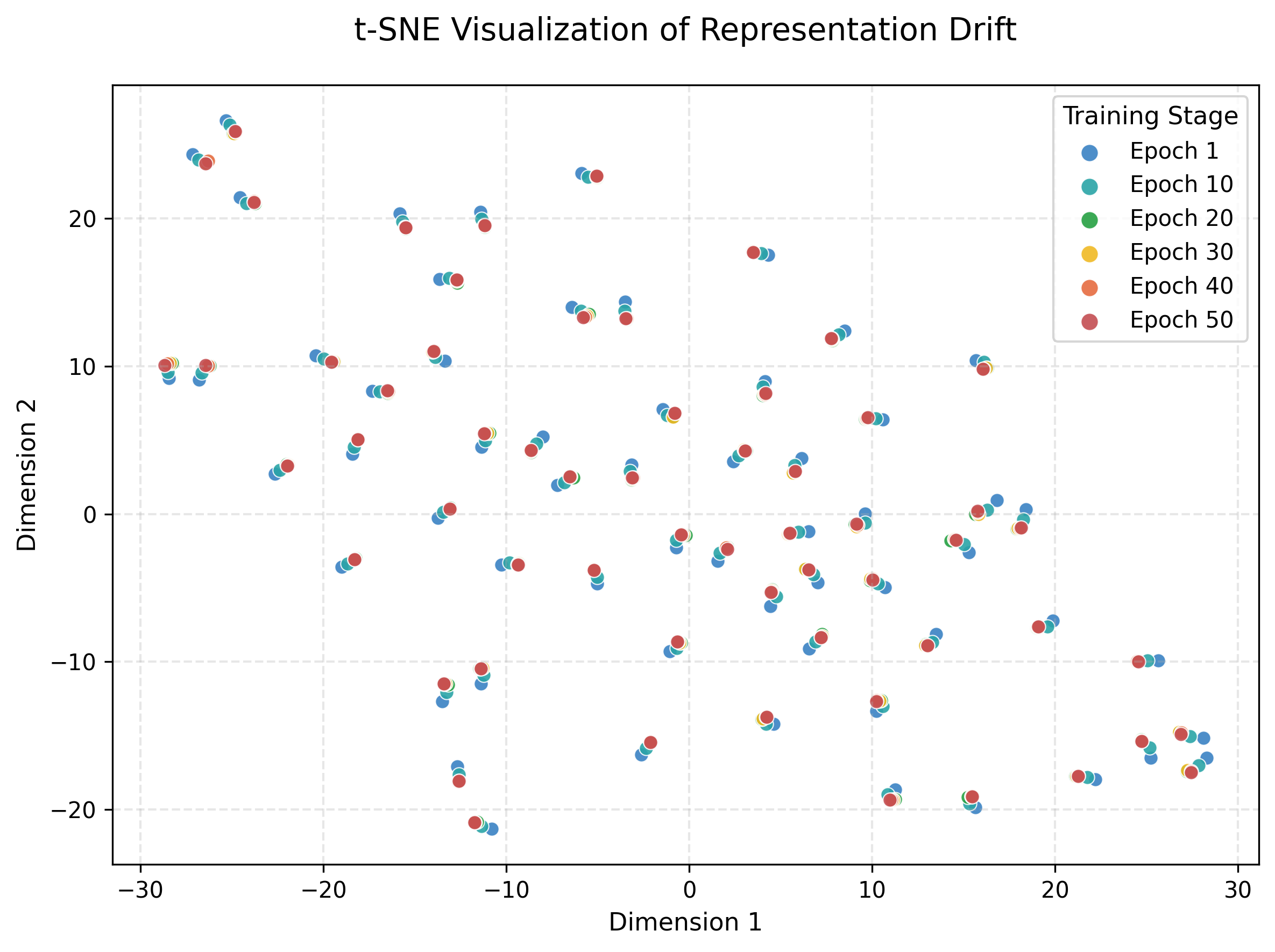}
  \caption{t-SNE visualization of Representation Drift. Embeddings of a fixed set of audio samples, encoded by the same model at different training epochs, are plotted. The progressive shift in embedding positions (from Epoch 1 [blue] to Epoch 50 [red]) empirically validates the core premise of RDM: a static knowledge base becomes misaligned with the non-stationary representation space over time.}
  \label{fig:tsne_drift}
\end{figure}

This appendix provides a detailed derivation of the relationship between the Representation Drift Mismatch (RDM) and training stability. The core premise of RDM is that a model's representation space is non-stationary during training. We first provide a visualization in Figure~\ref{fig:tsne_drift} that empirically demonstrates this phenomenon. It shows how the embeddings of the same audio clips, encoded by a model without dynamic updates, drift significantly as training progresses. Our goal in the following sections is to formally prove that this observed drift leads to a greater potential for gradient misalignment.

\paragraph{Gradient Formulation.}
We consider a simplified loss function $\mathcal{L} = \mathcal{L}_{\text{main}}(u_i, u'_i)$ that incorporates a knowledge-enhanced representation $u'_i = (1-\rho)u_i + \rho \mathcal{K}$, where $u_i = f_{\theta_t}(a_i)$ and $\mathcal{K}$ is the expected representation of retrieved knowledge. Formally, $\mathcal{K} = \sum_j P(j) z_j$, where $z_j$ denotes the embedding vector of the $j$-th sample in the knowledge base, and $P(j)$ is its corresponding probability weight. The gradient of the loss with respect to the model parameters $\theta_t$ is:
\begin{equation}
\label{eq:app_grad_base}
\nabla_{\theta_t} \mathcal{L} = \left( \frac{\partial \mathcal{L}}{\partial u_i} + (1-\rho) \frac{\partial \mathcal{L}}{\partial u'_i} \right) \frac{\partial u_i}{\partial \theta_t}
\end{equation}

\paragraph{Linking Gradient Deviation to Knowledge Deviation.}
The difference between the ideal gradient ($\nabla_{\theta_t} \mathcal{L}_{\text{ideal}}$) and the actual gradient ($\nabla_{\theta_t} \mathcal{L}_{\text{actual}}$) arises from the difference in their respective knowledge vectors, $\mathcal{K}_{\text{ideal}}$ and $\mathcal{K}_{\text{actual}}$. Let the gradient difference vector be $\Delta \nabla = \nabla_{\theta_t} \mathcal{L}_{\text{actual}} - \nabla_{\theta_t} \mathcal{L}_{\text{ideal}}$. This difference is primarily driven by the change in the loss derivative term $\frac{\partial \mathcal{L}}{\partial u'_i}$.

To analyze this relationship, we use a first-order Taylor expansion of the loss gradient term around the ideal representation $u'_{\text{ideal}}$. The difference can be approximated as:
\begin{equation}
\label{eq:app_taylor}
\frac{\partial \mathcal{L}_{\text{actual}}}{\partial u'_i} - \frac{\partial \mathcal{L}_{\text{ideal}}}{\partial u'_i} \approx H_{\mathcal{L}}(u'_{\text{ideal}}) \cdot (u'_{\text{actual}} - u'_{\text{ideal}})
\end{equation}
where $H_{\mathcal{L}}$ is the Hessian matrix of the loss function with respect to its input. Since $u'_{\text{actual}} - u'_{\text{ideal}} = \rho(\mathcal{K}_{\text{actual}} - \mathcal{K}_{\text{ideal}}) = \rho \Delta \mathcal{K}$, we can see that the deviation in the loss gradient is approximately proportional to the deviation in the knowledge vector:
\begin{equation}
\Delta \nabla \propto H_{\mathcal{L}} \cdot \Delta \mathcal{K}
\end{equation}
This establishes a direct relationship: a larger deviation in the fused knowledge vector $\Delta\mathcal{K}$ leads to a larger deviation in the final parameter gradient $\Delta \nabla$. The next step is therefore to bound the magnitude of $\Delta\mathcal{K}$ using the RDM.

\paragraph{Bounding the Knowledge Deviation via RDM}
We now bound the norm of the deviation $\|\Delta \mathcal{K}\|_2$ using the RDM. We leverage Pinsker's inequality, which relates the KL divergence to the Total Variation Distance ($D_{TV}$):
\begin{equation}
\begin{split}
D_{TV}(P_1, P_2) &= \frac{1}{2} \sum_{j} |P_1(j) - P_2(j)| \\
&\le \sqrt{\frac{1}{2} D_{KL}(P_1 \,||\, P_2)}
\end{split}
\end{equation}

Applying this to our distributions gives $D_{TV}(P_{\text{ideal}}, P_{\text{actual}}) \le \sqrt{\frac{1}{2} \text{RDM}(t, t_k)}$. We can then bound $\|\Delta \mathcal{K}\|_2$:
\begin{align}
\|\Delta &\mathcal{K}\|_2 = \| \sum_j (P_{\text{actual}}(j) - P_{\text{ideal}}(j)) z_j \|_2 \nonumber \\
&\le \sum_j |P_{\text{actual}}(j) - P_{\text{ideal}}(j)| \|z_j\|_2 \nonumber\\
&\le \left(\max_j \|z_j\|_2\right) \cdot 2 \cdot D_{TV}(P_{\text{ideal}}, P_{\text{actual}}) \nonumber \\
&\le C\sqrt{2 \cdot \text{RDM}(t, t_k)} \label{eq:app_final_bound}
\end{align}
where $C = \max_j \|z_j\|_2$ is a bounded constant.

\paragraph{Conclusion.}
Combining these steps, we have established a formal link: an increase in RDM widens the upper bound on the knowledge vector deviation $\|\Delta \mathcal{K}\|_2$ (Eq.~\ref{eq:app_final_bound}), which in turn increases the potential magnitude of the gradient deviation $\Delta \nabla$ (Eq.~\ref{eq:app_taylor}). This increases the risk of gradient misalignment, which can lead to training instability. Our dynamic knowledge refinement mechanism is designed to mitigate this risk by periodically resetting the RDM to zero.

\section{Theoretical Justification and Convergence of the ASK Objective}
\label{app:convergence}

In this section, we provide a complete theoretical justification for the ASK framework. We demonstrate that our training procedure can be viewed as a principled alternating optimization algorithm designed to maximize the log-likelihood of the observed data, which in turn guarantees the monotonic non-increase of our final loss function and thus ensures convergence.

\paragraph{Probabilistic Formulation with Latent Knowledge.}
The primary goal of Audio-Text Retrieval is to find model parameters $\theta^*$ that maximize the log-likelihood of observing matched audio-text pairs $x_i = (a_i, t_i)$:
\begin{equation}
\theta^* = \max_{\theta} \mathcal{L}(\theta) = \max_{\theta} \sum_i \log p(x_i; \theta)
\end{equation}
We conceptualize our approach by introducing latent variables, $z_i = (z_{i,f}, z_{i,c})$, representing the unobserved optimal knowledge for each sample $x_i$. The observed data likelihood is the marginal likelihood over these latent variables:
\begin{equation}
p(x_i; \theta) = \sum_{z_i} p(x_i, z_i; \theta)
\end{equation}
Thus, the optimization objective becomes:
\begin{equation}
\theta^* = \max_{\theta} \sum_i \log \sum_{z_i} p(x_i, z_i; \theta)
\end{equation}
The summation inside the logarithm makes direct optimization intractable.

\paragraph{Deriving the Evidence Lower Bound.}
To create a tractable objective, we introduce an arbitrary distribution $Q(z_i)$ and apply Jensen's Inequality to derive a lower bound on the log-likelihood, known as the Evidence Lower Bound (ELBO), denoted as $\mathcal{F}(Q, \theta)$:
\begin{align}
\log p(x_i; \theta) &= \log \sum_{z_i} Q(z_i) \frac{p(x_i, z_i; \theta)}{Q(z_i)} \nonumber \\
&\ge \sum_{z_i} Q(z_i) \log \frac{p(x_i, z_i; \theta)}{Q(z_i)} \label{eq:app_jensen} \\
\mathcal{F}(Q, \theta) &=\mathbb{E}_{Q(z_i)} [\log p(x_i, z_i; \theta)]\\
&\quad - \mathbb{E}_{Q(z_i)} [\log Q(z_i)] \nonumber
\end{align}
Maximizing $\log p(x_i; \theta)$ is achieved by iteratively maximizing this lower bound $\mathcal{F}$ with respect to $Q$ and $\theta$.

\paragraph{The ASK Framework as an Alternating Optimization Algorithm.}
Let $\theta_t$ be the parameters at iteration $t$. The ASK training process alternates between two stages.

\textbf{Stage 1: Auxiliary Distribution Update.}
In this stage, we fix $\theta_t$ and approximate the optimal auxiliary distribution $Q_t(z_i)$ which should be the true posterior $p(z_i|x_i;\theta_t)$. We assume independence between fine- and coarse-grained knowledge: $Q_t(z_i) = Q_{t,f}(z_{i,f})Q_{t,c}(z_{i,c})$.
\begin{itemize}
    \item The retrieval of Top-K neighbors defines the support of $Q_{t,f}$ and $Q_{t,c}$.
    \item We define the probability mass of these distributions over a specific neighbor $z_j$ using our reliability weights:
    \begin{equation}
    \begin{split}
    \label{eq:app_q_def}
    Q_{t,f}(z_{i,f}=z_j) &:= w_{j,f}(\theta_t), \\
    Q_{t,c}(z_{i,c}=z_j) &:= w_{j,c}(\theta_t)
    \end{split}
    \end{equation}
\end{itemize}

\textbf{Stage 2: Model Parameter Update.}
In this stage, we fix $Q_t$ and maximize the ELBO with respect to $\theta$, which is equivalent to maximizing $\mathbb{E}_{Q_t} [\log p(x_i, z_i; \theta)]$. We model the joint log-probability as a sum of independent fine- and coarse-grained components, e.g., for the text-to-audio direction:
\begin{equation}
\label{eq:app_logp_model}
\begin{split}
\log p(x_i, z_i; \theta) \approx  &\left( -\mathcal{L}_{OT, f}(\theta) - \log \Psi_{i,f}^{T \leftarrow A}(\theta) \right) \\
&+ \left( -\mathcal{L}_{OT, c}(\theta) - \log \Psi_{i,c}^{T \leftarrow A}(\theta) \right)\\
&+ \left(-\log Z(\theta)\right)
\end{split}
\end{equation}
where $Z(\theta)$ is a normalization constant. The maximization objective is to minimize the negative expectation of this log-probability under $Q_t$. Substituting Eq.~\ref{eq:app_q_def} and Eq.~\ref{eq:app_logp_model}, this objective becomes:
\begin{equation}
\begin{split}
\mathcal{L}_{m} &= -\sum_i \mathbb{E}_{Q_t(z_i)} [\log p(x_i, z_i; \theta)] \\
&\approx \sum_i ( \mathbb{E}_{Q_{t,f}}[\mathcal{L}_{OT, f} + \log \Psi_{i,f}] \\
&\quad + \mathbb{E}_{Q_{t,c}}[\mathcal{L}_{OT, c} + \log \Psi_{i,c}] )
\end{split}
\end{equation}
Our final modulated loss,
\begin{equation}
\mathcal{L}^*_{T \to A} = (1 + \lambda_f \mathcal{F}_{f}^{T \to A} + \lambda_c \mathcal{F}_{c}^{T \to A}) \cdot \mathcal{L}_{T \to A}
\end{equation}
where $\mathcal{F} = -\log \Psi$, is a principled and sophisticated implementation of this maximization objective. Minimizing $\mathcal{L}_{\text{ASK}}$ effectively performs this parameter update.

\paragraph{Proof of Convergence.}
This two-stage alternating optimization guarantees that the total objective is non-decreasing at each full iteration, $\mathcal{L}(\theta_{t+1}) \ge \mathcal{L}(\theta_t)$. Consequently, minimizing the negative log-likelihood guarantees that the loss is monotonically non-increasing. Given that $\mathcal{L}_{\text{ASK}}$ is bounded below by zero, the Monotone Convergence Theorem ensures that the sequence of loss values converges to a limit, and the parameters $\{\theta_t\}$ converge to a stationary point

\section{Optimal Transport for Batch-level Alignment}
\label{app:ot_details}

This section details the entropy-regularized Optimal Transport (OT) formulation used to refine the batch-wise similarity matrices. Given a batch of knowledge-enhanced pairs, we compute a similarity matrix, e.g., the fine-grained matrix $\mathbf{S}_f \in \mathbb{R}^{B\times B}$. We then seek an optimal transport plan $\mathbf{Q} \in \mathbb{R}^{B\times B}$, where $\mathbf{Q}_{ij}$ represents the soft-alignment probability between the $i$-th text and the $j$-th audio. The optimal plan $\mathbf{Q}^*$ is found by solving the following regularized optimization problem:
\begin{equation}
\begin{aligned}
\mathbf{Q}^* = \max_{\mathbf{Q}\in\mathcal{C}}\  &\langle \mathbf{Q}, \mathbf{S}_f\rangle + \varepsilon H(\mathbf{Q})\\
\text{s.t.} \ \mathcal{C}=\{\mathbf{Q}\in \mathbb{R}^{B\times B}&\mid \mathbf{Q}\mathbf{1}_{B}=\boldsymbol{\mu},\ \mathbf{Q}^\top\mathbf{1}_{B}=\boldsymbol{\nu}\},
\end{aligned}
\end{equation}
where $\langle\mathbf{Q},\mathbf{S}_f\rangle = \mathrm{tr}(\mathbf{Q}^\top \mathbf{S}_f)$ is the total similarity score. $H(\mathbf{Q}) = -\sum_{i,j}\mathbf{Q}_{ij}\log\mathbf{Q}_{ij}$ is the entropy regularizer, controlled by $\varepsilon > 0$. The constraints enforce that the marginals of $\mathbf{Q}$ must sum to predefined distributions $\boldsymbol{\mu}$ and $\boldsymbol{\nu}$, which represent the importance of each instance. Following prior work \citep{su2017order}, we set both $\boldsymbol{\mu}$ and $\boldsymbol{\nu}$ to a uniform distribution over the batch, i.e., $\frac{1}{|B|}\mathbf{1}_{|B|}$. This problem is efficiently solved for the optimal plan $\mathbf{Q}^*$ using the Sinkhorn-Knopp algorithm \citep{cuturi2013sinkhorn}.

\section{Full Results for Local Interaction Strategy}
\label{app:local_results}

This section provides the complete retrieval results for our experiments on the local, token-level interaction baselines, including both Audio-to-Text and Text-to-Audio directions. Table~\ref{tab:full_local_results} presents the full comparison against both the FLAM and GPA architectures.

\begin{table}[ht!]
    \centering
    \caption{Full results for Audio-Text Retrieval on AudioCaps and Clotho under the local interaction strategy. The symbols $^+$, $^\dagger$, and $^*$ denote different knowledge sources.} 
    \label{tab:full_local_results}
    \resizebox{\columnwidth}{!}{%
    \begin{tabular}{l|ccc|ccc}
        \toprule
        \multicolumn{7}{c}{\textbf{Audio-to-Text}} \\
        \cmidrule(lr){1-7}
        & \multicolumn{3}{c|}{AudioCaps} & \multicolumn{3}{c}{Clotho} \\
        \cmidrule(lr){1-7}
        Method & R@1 & R@5 & R@10 & R@1 & R@5 & R@10 \\
        \cmidrule(lr){1-7}
        FLAM \cite{wu2025flam} & $40.3_{\scriptsize \pm 0.1}$ & $71.2_{\scriptsize \pm 0.3}$ & $83.6_{\scriptsize \pm 0.2}$ & $17.5_{\scriptsize \pm 0.1}$ & $38.8_{\scriptsize \pm 0.4}$ & $50.5_{\scriptsize \pm 0.4}$ \\
        GPA \cite{xie2024gpa} & $41.1_{\scriptsize \pm 0.3}$ & $73.8_{\scriptsize \pm 0.4}$ & $85.2_{\scriptsize \pm 0.6}$ & $18.1_{\scriptsize \pm 0.2}$ & $40.2_{\scriptsize \pm 0.3}$ & $53.4_{\scriptsize \pm 0.4}$ \\
        \textbf{ASK}$^\dagger$    & $42.9_{\scriptsize \pm 0.3}$ & $75.1_{\scriptsize \pm 0.6}$ & $86.4_{\scriptsize \pm 0.5}$ & $19.1_{\scriptsize \pm 0.1}$ & $\textbf{41.9}_{\scriptsize \pm 0.4}$ & $53.9_{\scriptsize \pm 0.8}$  \\
        \textbf{ASK}$^*$          & $\textbf{43.7}_{\scriptsize \pm 0.2}$ & $\textbf{75.8}_{\scriptsize \pm 0.3}$ & $86.2_{\scriptsize \pm 0.7}$ & $19.2_{\scriptsize \pm 0.3}$ & $41.6_{\scriptsize \pm 0.7}$ & $\textbf{54.5}_{\scriptsize \pm 0.6}$ \\
        \textbf{ASK}$^+$          & $43.1_{\scriptsize \pm 0.3}$ & $74.0_{\scriptsize \pm 0.6}$ & $\textbf{86.9}_{\scriptsize \pm 0.5}$ & $\textbf{19.5}_{\scriptsize \pm 0.3}$ & $41.4_{\scriptsize \pm 0.7}$ & $\textbf{54.5}_{\scriptsize \pm 0.6}$ \\
        \midrule
        \multicolumn{7}{c}{\textbf{Text-to-Audio}} \\
        \cmidrule(lr){1-7}
        & \multicolumn{3}{c|}{AudioCaps} & \multicolumn{3}{c}{Clotho} \\
        \cmidrule(lr){1-7}
        Method & R@1 & R@5 & R@10 & R@1 & R@5 & R@10 \\
        \cmidrule(lr){1-7}
        FLAM \cite{wu2025flam}    & $33.1_{\scriptsize \pm 0.1}$ & $67.5_{\scriptsize \pm 0.2}$ & $80.0_{\scriptsize \pm 0.2}$ & $13.8_{\scriptsize \pm 0.2}$ & $33.2_{\scriptsize \pm 0.3}$ & $45.1_{\scriptsize \pm 0.2}$ \\
        GPA \cite{xie2024gpa}    & $34.1_{\scriptsize \pm 0.2}$ & $70.0_{\scriptsize \pm 0.4}$ & $82.2_{\scriptsize \pm 0.6}$ & $15.1_{\scriptsize \pm 0.2}$ & $37.9_{\scriptsize \pm 0.6}$ & $50.2_{\scriptsize \pm 0.4}$ \\
        \textbf{ASK}$^\dagger$  & $34.5_{\scriptsize \pm 0.3}$ & $\textbf{71.1}_{\scriptsize \pm 0.5}$ & $\textbf{83.1}_{\scriptsize \pm 0.6}$ & $16.2_{\scriptsize \pm 0.1}$ & $38.5_{\scriptsize \pm 0.4}$ & $51.3_{\scriptsize \pm 0.5}$\\
        \textbf{ASK}$^*$        & $34.6_{\scriptsize \pm 0.2}$ & $70.5_{\scriptsize \pm 0.5}$ & $82.7_{\scriptsize \pm 0.6}$ & $\textbf{16.3}_{\scriptsize \pm 0.2}$ & $38.4_{\scriptsize \pm 0.3}$ & $51.5_{\scriptsize \pm 0.4}$ \\
        \textbf{ASK}$^+$        & $\textbf{35.1}_{\scriptsize \pm 0.3}$ & $70.8_{\scriptsize \pm 0.5}$ & $\textbf{83.1}_{\scriptsize \pm 0.4}$ & $16.0_{\scriptsize \pm 0.1}$ & $\textbf{38.8}_{\scriptsize \pm 0.3}$ & $\textbf{52.1}_{\scriptsize \pm 0.5}$\\
        \bottomrule
    \end{tabular}
    } 
\end{table}

As demonstrated in Table~\ref{tab:full_local_results}, ASK consistently outpaces both baselines across all metrics in the Text-to-Audio retrieval direction as well. When compared to the stronger GPA baseline, ASK$^+$ achieves the highest R@1 score on AudioCaps, yielding a 1.0\% absolute improvement, which translates to a substantial 2.0\% margin over FLAM. On Clotho, the ASK$^*$ variant delivers the strongest R@1 performance with a significant gain of 1.2\% absolute over GPA, and a notable 2.5\% absolute advantage over FLAM. These results confirm that the benefits of our proposed mechanisms are symmetric, enhancing both retrieval directions and validating the overall effectiveness of the ASK framework on fine-grained architectures.

\section{Comparison with Alternative Retrieval-Augmented Frameworks}

To further validate the design choices of our ASK framework, we compare it against classical retrieval-augmented contrastive learning paradigms adapted from computer vision and information retrieval. Specifically, we benchmark against representative retrieval-augmented and contrastive learning paradigms: MoCo \cite{he2020momentum}, ANCE \cite{xiong2020approximate}, and NNCLR \cite{dwibedi2021little}. These baselines represent the classical strategies for expanding negative sample capacity, performing hard negative mining, and utilizing external support sets, respectively. We adapt their core mechanisms to the Audio-Text Retrieval setting. All variants are built upon the identical ML-ACT \cite{mei2022metric} backbone to ensure a fair and rigorous comparison.

\begin{table}[ht!]
    \centering
    \caption{Comparison with other retrieval-augmented methods on AudioCaps. The symbol $+$ denotes using training set as the knowledge source} 
    \label{tab:compare_results}
    \resizebox{\columnwidth}{!}{%
    \begin{tabular}{l|ccc|ccc}
        \toprule
        & \multicolumn{3}{c|}{Audio-to-Text} & \multicolumn{3}{c}{Text-to-Audio} \\
        \cmidrule(lr){1-7}
        Method & R@1 & R@5 & R@10 & R@1 & R@5 & R@10 \\
        \cmidrule(lr){1-7}
        ML-ACT \cite{mei2022metric} & $36.3_{\pm 0.5}$ & $68.6_{\pm 0.3}$ & $81.5_{\pm 0.2}$ & $32.2_{\pm 0.4}$ & $68.2_{\pm 0.1}$ & $81.2_{\pm 0.2}$\\
        NNCLR \cite{dwibedi2021little} & $32.5_{\pm 0.9}$ & $66.6_{\pm 0.5}$ & $80.9_{\pm 0.7}$ & $30.2_{\pm 0.3}$ & $64.9_{\pm 0.1}$ & $80.1_{\pm 0.1}$\\
        ANCE \cite{xiong2020approximate} & $35.9_{\scriptsize \pm 0.5}$ & $67.8_{\scriptsize \pm 0.3}$ & $83.6_{\scriptsize \pm 0.2}$ & $31.7_{\scriptsize \pm 0.1}$ & $65.4_{\scriptsize \pm 0.5}$ & $79.8_{\scriptsize \pm 0.1}$ \\
        MoCo \cite{he2020momentum} & $37.2_{\scriptsize \pm 0.3}$ & $68.5_{\scriptsize \pm 0.5}$ & $82.0_{\scriptsize \pm 0.3}$ & $32.9_{\scriptsize \pm 0.4}$ & $67.4_{\scriptsize \pm 0.6}$ & $80.9_{\scriptsize \pm 0.1}$ \\
        \textbf{ASK}$^+$        & $\textbf{42.0}_{  \pm 0.2}$ & $\textbf{74.2}_{  \pm 0.5}$ & $\textbf{85.4}_{  \pm 0.6}$ & $\textbf{35.4}_{  \pm 0.3}$ & $\textbf{70.2}_{  \pm 0.3}$ & $\textbf{83.1}_{  \pm 0.7}$\\
        \bottomrule
    \end{tabular}
    } 
\end{table}

As shown in Table \ref{tab:compare_results}, our proposed \textbf{ASK} significantly outperforms all alternative strategies across all metrics, achieving an R@1 of 42.0\% in Audio-to-Text retrieval. Conversely, directly migrating classical contrastive methods to the Audio-Text Retrieval (ATR) domain yields sub-optimal or degraded performance. While MoCo provides only marginal improvements, both ANCE and NNCLR lead to noticeable performance drops, with NNCLR causing the most severe degradation (e.g., Audio-to-Text R@1 dropping from 36.3\% to 32.5\%).

We hypothesize that this performance degradation stems from the unique semantic ambiguities in ATR. Specifically, methods relying on strict hard negative mining (ANCE) or unimodal nearest neighbors (NNCLR) are highly susceptible to acoustic confusion, which inadvertently introduces false negatives or modality-specific noise into the contrastive objective. Additionally, the memory queues in MoCo may suffer from representation drift due to the continuous updating of the fine-grained audio encoder. Unlike these rigid mechanisms, ASK mitigates these domain-specific bottlenecks by employing an adaptive reliability weighting scheme that explicitly evaluates cross-modal consistency, effectively filtering out the retrieval noise that limits standard baselines.

\section{Hyperparameter Sensitivity Analysis}
\label{sec:hyperparameter_analysis}

\begin{figure}[htbp]
  \centering
  \begin{subfigure}{0.48\columnwidth}
    \centering
    \includegraphics[width=\textwidth]{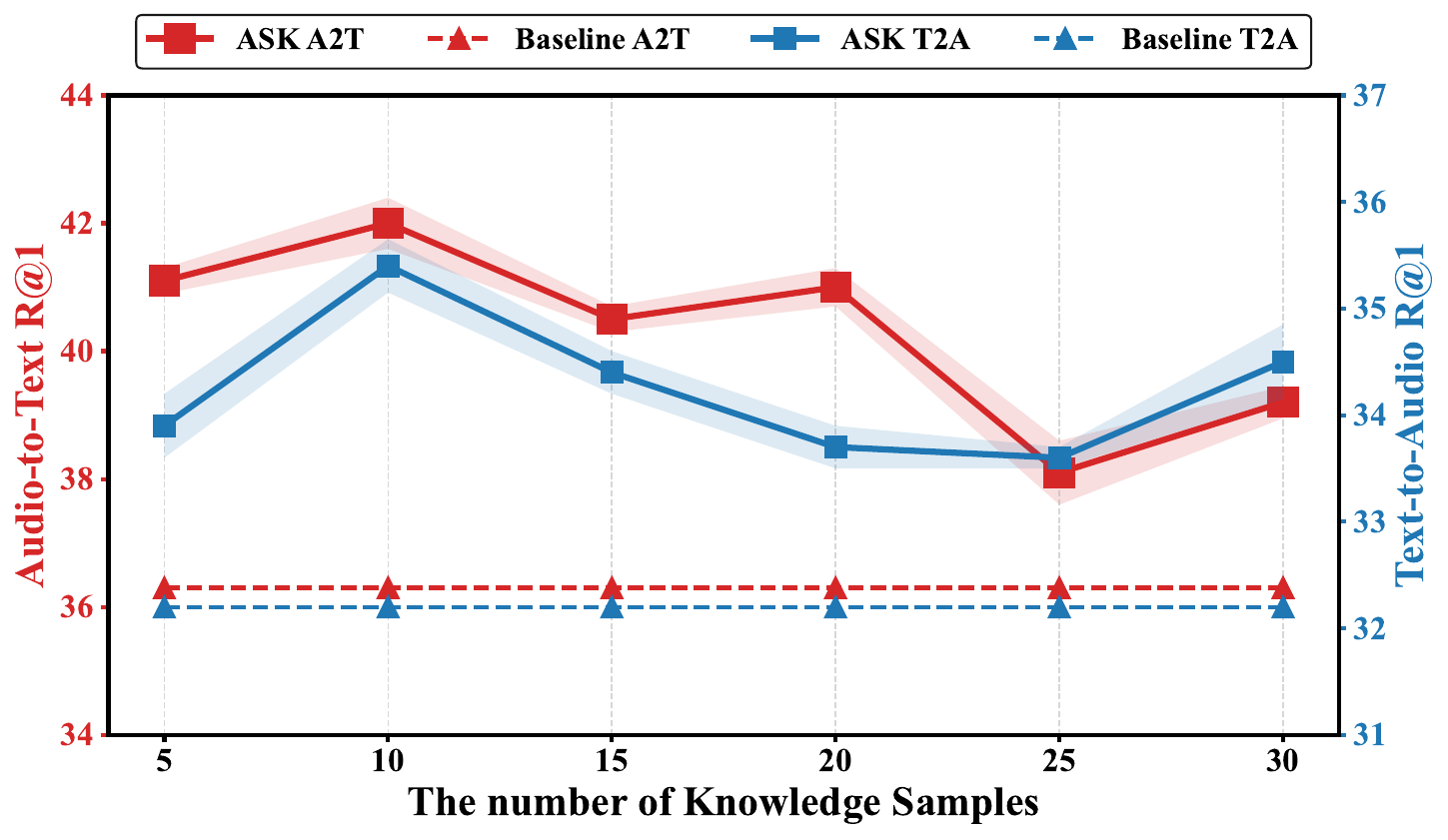}
  \end{subfigure}
  \hfill 
  \begin{subfigure}{0.48\columnwidth}
    \centering
    \includegraphics[width=\textwidth]{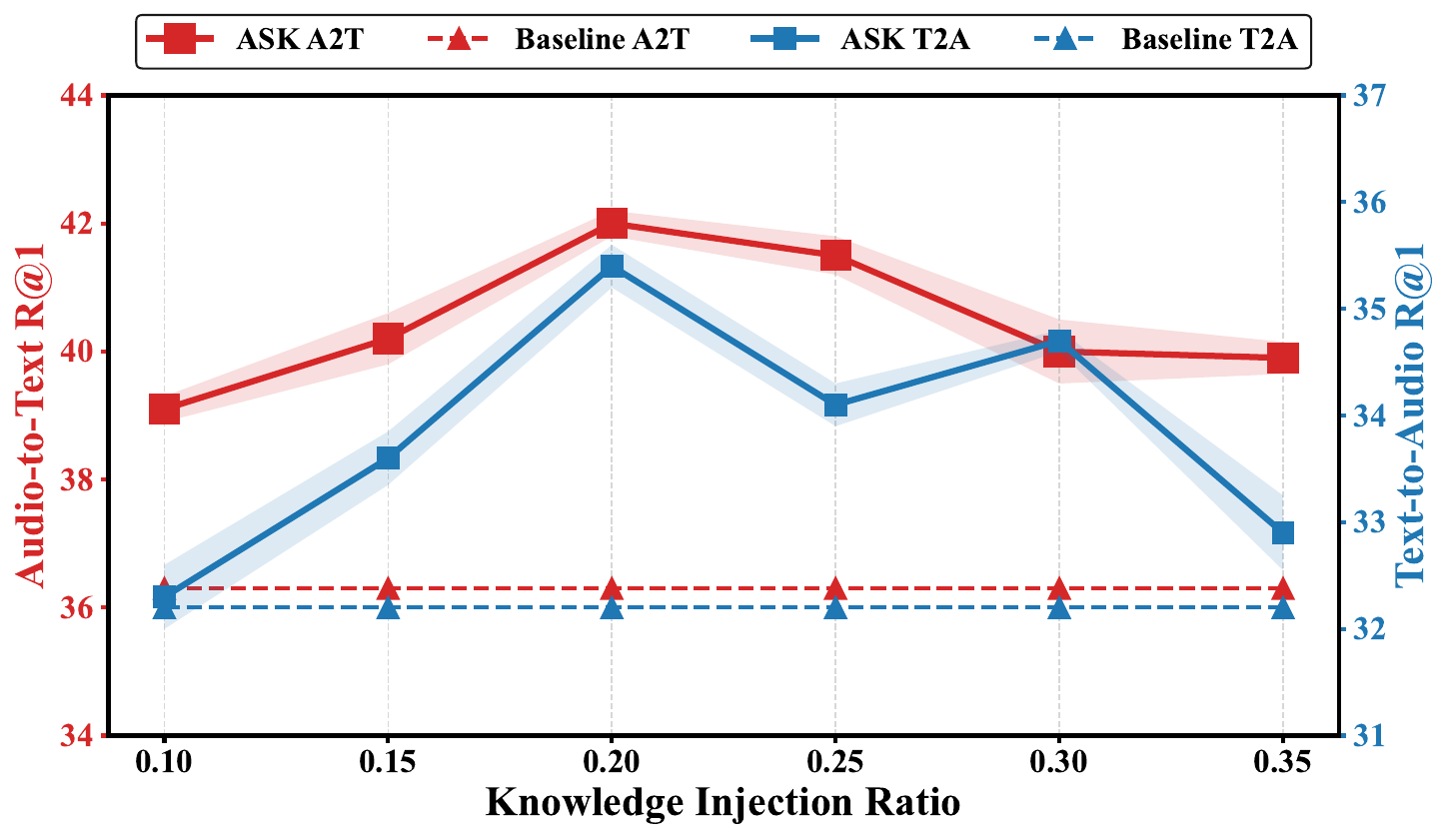}
  \end{subfigure}
  
  \caption{Hyperparameter sensitivity analysis. Left: Impact of the retrieved knowledge size $K$. Right: Impact of the knowledge injection ratio $\rho$.}
  \label{fig:hyperparameters}
\end{figure}

To comprehensively evaluate the robustness of the proposed ASK framework, we conduct sensitivity analyses on two critical hyperparameters: the retrieved knowledge size ($K$) and the knowledge injection ratio ($\rho$). The experiments are conducted on the AudioCaps dataset using the ASK$^+$ variant. The results are illustrated in Figure \ref{fig:hyperparameters}.

\paragraph{\textbf{Impact of the Retrieved Knowledge Size ($K$).}} The left panel of Figure \ref{fig:hyperparameters} illustrates the effect of varying the number of retrieved knowledge samples $K$ from $5$ to $30$. The framework achieves its optimal performance in both Audio-to-Text (A2T) and Text-to-Audio (T2A) retrieval at $K=10$, reaching a peak A2T R@1 of 42.0\% and T2A R@1 of 35.4\%. When $K$ is set too small (e.g., $K=5$), the limited neighborhood fails to provide sufficient semantic diversity, restricting the model's ability to generalize to ambiguous or long-tail concepts. Conversely, setting $K$ too large (e.g., $K \ge 15$) inevitably introduces acoustically similar but semantically irrelevant noise into the knowledge base. This excessive retrieval dilutes the valid semantic guidance and consequently degrades the alignment performance.

\paragraph{\textbf{Impact of the Knowledge Injection Ratio ($\rho$).}} The right panel of Figure \ref{fig:hyperparameters} presents the performance fluctuations across different knowledge injection ratios $\rho \in [0.10, 0.35]$. This hyperparameter controls the crucial trade-off between retaining the original instance identity and incorporating the retrieved global semantic prior. As shown in the figure, the model attains its peak performance at $\rho=0.20$. A lower injection ratio (e.g., $\rho=0.10$) provides insufficient global context, making it difficult to fully break the Gradient Locality Bottleneck (GLB). However, an excessively high ratio ($\rho \ge 0.25$) over-dominates the feature fusion process. This leads to feature homogenization, where the unique acoustic or textual characteristics of the original sample are overshadowed by the aggregated neighborhood representations, ultimately resulting in a decline in retrieval accuracy.

\section{Visualization of Adaptive Reliability Weighting}
\label{sec:weight_visualization}

To explicitly address how the Adaptive Reliability Weighting mechanism mitigates Representation-Drift Mismatch (RDM) and filters noise, we delve into the mathematical behavior of the reliability potential $\mathcal{F} = -\log \Psi$. We provide an internal visualization of these assigned weights during the training process to demonstrate their dynamic modulation effects.

A critical but nuanced property of our framework is that $\mathcal{F}$ acts as a dynamic negative regularizer. Given the normalized embedding space, the cross-modal similarity generally maintains non-negative exponential characteristics, leading to an expected similarity $\Psi \ge 1$. Consequently, the reliability weight $\mathcal{F}$ consistently operates in the negative domain ($\mathcal{F} \le 0$). This design intrinsically serves as a reward mechanism that dynamically relaxes the standard in-batch contrastive penalty when reliable external knowledge is injected.

Table \ref{tab:weight_vis} illustrates this internal mechanism using two challenging acoustic ambiguity scenarios. Query 1 involves confusing environmental textures (\textit{``Thunder''} vs. \textit{``Sizzling food''}). Query 2, drawn directly from our evaluation set, involves engine and wind noises where a \textit{``Motorboat''} could be acoustically confused with a \textit{``Motorcycle''}.

\begin{table}[htbp]
\centering
\caption{Internal visualization of the reliability potential $\Psi$ and the negative regularizer weight $\mathcal{F}$.}
\label{tab:weight_vis}
\resizebox{\linewidth}{!}{%
\begin{tabular}{llccc}
\toprule
\textbf{Neighborhood State} & \textbf{Semantic Concept} & \textbf{$\Psi$} & \textbf{$\mathcal{F}$} & \textbf{Modulation} \\
\midrule
\multicolumn{5}{l}{\textbf{Query 1:} \textit{``Thunder roars in the distance as rain falls''} (Environmental Texture Ambiguity)} \\
\midrule
\textbf{Clean Neighborhood} & \textit{``Loud thunder claps''} & \textbf{1.60} & \textcolor{green!60!black}{\textbf{- 0.47}} & \textbf{0.91} \\
\addlinespace
\textbf{Noise Neighborhood} & \textit{``Food sizzling loudly''} & \textbf{1.05} & \textcolor{red!80!black}{\textbf{- 0.05}} & \textbf{0.99}\\
\midrule
\multicolumn{5}{l}{\textbf{Query 2:} \textit{``A motorboat driving by as water splashes and wind blows''} (Vehicle Engine Ambiguity)} \\
\midrule
\textbf{Clean Neighborhood} & \textit{``A speedboat traveling across water''} & \textbf{1.55} & \textcolor{green!60!black}{\textbf{- 0.44}} & \textbf{0.91}\\
\addlinespace
\textbf{Noise Neighborhood} & \textit{``A motorcycle engine revving''} & \textbf{1.08} & \textcolor{red!80!black}{\textbf{- 0.08}} & \textbf{0.98}\\
\bottomrule
\end{tabular}%
}
\end{table}

When the retrieved neighborhood contains highly relevant concepts (e.g., \textit{``Loud thunder claps''} or \textit{``A speedboat traveling''}), the strong semantic alignment yields a larger $\Psi$ ($>1$), forcing $\mathcal{F}$ to be a distinctively negative value with a large magnitude (e.g., -0.47 and -0.44). This dynamically reduces the modulation multiplier $(1 + \lambda \mathcal{F}) < 1$, effectively rewarding the model by down-weighting the standard contrastive penalty and encouraging the absorption of this clean knowledge. 

Conversely, when the retrieved knowledge is corrupted by RDM drift—such as retrieving the acoustically similar but semantically incorrect \textit{``Food sizzling loudly''} or \textit{``A motorcycle engine revving''}—the weak cross-modal alignment results in a $\Psi$ closer to 1. This causes the negative regularizer $\mathcal{F}$ to approach zero (e.g., -0.05 and -0.08), which pulls the modulation multiplier back to near 1.0. By doing so, the framework withholds the reward and maintains the strict original contrastive loss, seamlessly preventing the model from over-relying on the drifted acoustic noise.

\end{document}